\documentclass[pdflatex,sn-mathphys-num]{sn-jnl}

\usepackage{graphicx}
\usepackage{colortbl}
\usepackage{booktabs}
\usepackage{multirow}
\usepackage{array}
\usepackage{amsmath,amssymb,amsfonts}
\usepackage{amsthm}
\usepackage{mathrsfs}
\usepackage[title]{appendix}
\usepackage{textcomp}
\usepackage{manyfoot}
\usepackage{booktabs}
\usepackage{algorithm}
\usepackage{algorithmicx}
\usepackage{algpseudocode}
\usepackage{listings}
\usepackage{caption}
\usepackage[nonumberlist]{glossaries}
\usepackage{placeins} 
\usepackage[table]{xcolor}
\glsdisablehyper

\usepackage{xcolor}
\definecolor{slice1}{HTML}{4c72b0} 
\definecolor{slice2}{HTML}{55a868} 
\definecolor{slice3}{HTML}{c44e52} 
\definecolor{slice4}{HTML}{8172b2} 
\definecolor{slice5}{HTML}{ccb974} 
\definecolor{others}{HTML}{ff7f0e} 

\newacronym{sms}{SMS}{\textit{Short Message Service}}
\newacronym{mms}{MMS}{\textit{Multimedia Messaging Services}}
\newacronym{qos}{QoS}{\textit{Quality of Service}}
\newacronym{sba}{SBA}{\textit{Service-based Architecture}}
\newacronym{amf}{AMF}{\textit{Access and Mobility Management Function}}
\newacronym{smf}{SMF}{\textit{Session Management Function}}
\newacronym{udm}{UDM}{\textit{Unified Data Management}}
\newacronym{udr}{UDR}{\textit{Unified Data Repository}}
\newacronym{nssf}{NSSF}{\textit{Network Slice Selection Function}}
\newacronym{scp}{SCP}{\textit{Service Communication Proxy}}
\newacronym{ausf}{AUSF}{\textit{Authentication Serve Function}}
\newacronym{pcf}{PCF}{\textit{Policy Control Function}}
\newacronym{upf}{UPF}{\textit{User Plane Function}}
\newacronym{nrf}{NRF}{\textit{Network Repository Function}}
\newacronym{nf}{NF}{\textit{Network Function}}
\newacronym{bsf}{BSF}{\textit{Binding Support Function}}
\newacronym{sepp}{SEPP}{\textit{Security Edge Protection Proxy}}
\newacronym{ue}{UE}{\textit{User Equipment}}
\newacronym{ran}{RAN}{\textit{Radio Access Network}}
\newacronym{nfv}{NFV}{\textit{Network Function Virtualization}}
\newacronym{sdn}{SDN}{\textit{Software-Defined Networking}}
\newacronym{cots}{COTS}{\textit{Commercial Off-The-Shelf}}
\newacronym{up}{UP}{\textit{User Plane}}
\newacronym{cp}{CP}{\textit{Control Plane}}
\newacronym{nssai}{NSSAI}{\textit{Network Slice Selection Assistance Information}}
\newacronym{s-nssai}{S-NSSAI}{\textit{Single – Network Slice Selection Assistance Information}}
\newacronym{sst}{SST}{\textit{Slice/Service Type}}
\newacronym{sd}{SD}{\textit{Slice Differentiator}}
\newacronym{mano}{MANO}{\textit{Management and Orchestration}}
\newacronym{cncf}{CNCF}{\textit{Cloud Native Computing Foundation}}
\newacronym{iac}{IaC}{Infrastructure as Code}
\newacronym{lb}{LB}{\textit{Local Breakout}}
\newacronym{iqr}{IQR}{Intervalo Interquartil}
\newacronym{e2e}{E2E}{\textit{End-to-end}}
\newacronym{gnb}{gNB}{\textit{gNodeB}}
\newacronym{3gpp}{3GPP}{\textit{3rd Generation Partnership Project}}
\newacronym{vnf}{VNF}{\textit{Virtual Network Function}}
\newacronym{ric}{RIC}{\textit{RAN Intelligent Controller}}
\newacronym{embb}{eMBB}{\textit{Enhanced Mobile Broadband}}
\newacronym{urllc}{uRLLC}{\textit{Ultra Reliable Low Latency Communications}}
\newacronym{mmtc}{mMTC}{\textit{Massive Machine-Type Communications}}
\newacronym{iot}{IoT}{\textit{Internet of Things}}
\newacronym{5g}{5G}{Fifth Generation of Mobile Networks}
\newacronym{5gc}{5GC}{5G Network Core}
\newacronym{nsis}{NSIs}{\textit{Network Slice Instances}}
\newacronym{nsi}{NSI}{\textit{Network Slice Instance}}
\newacronym{cni}{CNI}{\textit{Container Network Interface}}

\theoremstyle{thmstyleone}%
\theoremstyle{thmstyletwo}%

\theoremstyle{thmstylethree}%

\begin{document}

\title[Article Title]{A Study on 5G Network Slice Isolation Based on Native Cloud and Edge Computing Tools}


\author*[1]{\fnm{Maiko} \sur{Andrade}}\email{maikovisky@gmail.com}

\author[1]{\fnm{Juliano} \sur{Wickboldt}}\email{jwickboldt@inf.ufrgs.br}

\affil*[1]{\orgdiv{Federal University of Rio Grande do Sul}, \orgname{UFRGS}, \orgaddress{\street{Av. Bento Gonçalves 9500}, \city{Porto Alegre}, \postcode{90046-900}, \state{RS}, \country{Brazil}}}


\abstract{5G networks support various advanced applications through network slicing, network function virtualization (NFV), and edge computing, ensuring low latency and service isolation. However, private 5G networks relying on open-source tools still face challenges in maturity and integration with edge/cloud platforms, compromising proper slice isolation. This study investigates resource allocation mechanisms to address this issue, conducting experiments in a hospital scenario with medical video conferencing. The results show that CPU limitations improve the performance of prioritized slices, while memory restrictions have minimal impact. The generated data and scripts have been made publicly available for future research and machine learning applications.}




\maketitle

\section{Introduction}\label{sec1}


\gls{5g} explore various advanced applications and technologies, which enabled innovation in sectors such as entertainment, virtual reality, tactile Internet, the Internet of Things, intelligent transportation systems, and emergency response~\cite{Hassan2019}. In the field of entertainment, these technologies enable the transmission of high-definition content. Virtual, augmented, and mixed realities are becoming more accessible through the use of edge computing. The tactile Internet relies on ultra-responsive networks for critical applications. Additionally, the Internet of Things connects home devices, while intelligent transportation systems improve road safety and the operation of autonomous vehicles. Emergency response is enhanced with real-time data collection, enabling remote surgeries and diagnoses.

These are examples of applications that fit within the three broad scenarios for which \gls{5g} is designed:  \gls{embb}, \gls{urllc}, and \gls{mmtc} \cite{Farooqui:2022}. To meet these scenarios, network slicing was introduced, which allows multiple logical networks or instances to operate simultaneously on the same common infrastructure. It is essential to ensure the isolation of each network instance so that one slice does not interfere with another, avoiding performance degradation, failures, or security breaches. This is crucial for maintaining service guarantees and facilitating root cause identification for issues in a \gls{nsi} \cite{Gonzalez:2020}.

In addition to the performance improvements brought by \gls{5g} networks with the introduction of slicing, the advantages of \gls{sba} are leveraged. This architecture has brought the worlds of telecommunications and information technology closer together, moving away from approaches based on purpose-specific equipment and favoring \gls{cots} solutions \cite{Luong2018}. 
Network function virtualization (NFV), in turn, enables network functions to run in virtualized environments (\textit{e.g.}, virtual machines or containers), promoting scalable and automated networks \cite{Hassan2019}. All these innovations allow the implementation of 5G infrastructure in an agile and cost-effective manner through the use of \gls{vnf}, in addition to bringing benefits to network operators \cite{Luong2018}.

Another advantage of using \gls{nfv} is the possibility of physically bringing the infrastructure closer to users' mobile devices, or \gls{ue}, through edge computing. Unlike centralized cloud computing, which relies on distant data centers, edge computing meets the needs of applications requiring low latency and high throughput. By processing data locally, it also reduces operational costs by avoiding the transmission of tasks and data to the cloud, thereby decreasing bandwidth consumption and other network resources \cite{Hassan2019}. On the flip side of edge computing, resource management, and in particular slice management, becomes more challenging due to the resource-constrained devices at the edge. The limited processing power, memory, and energy resources available at the edge demand efficient allocation strategies to maintain performance while ensuring isolation across network slices.

Due to the adoption of modern concepts such as \gls{sba} and \gls{nfv}, \gls{5g} offers the real possibility of creating private networks, allowing for the implementation of smaller-scale networks in locations such as university campuses, testbed environments, hospital complexes, or factories. Unlike national operator networks, which acquire complete and high-cost solutions, private network deployments may not always have required resources and therefore turn to open-source tools, whose implementations still have a reduced level of maturity for production networks \cite{Silveira2022, Lando2023}. These private networks must support slicing to ensure the proper functioning of various applications and their requirements. Despite open-source implementations natively supporting slice configuration, the actual integration with edge/cloud infrastructures, based for instance on Kubernetes, may not provide the adequate and necessary isolation to meet application requirements. Consequently, when needed, the operator may lack effective means to prioritize one network slice over others.

Considering the problem of the immaturity of open-source tools for implementing private \gls{5g} networks and their lack of integration with edge/cloud platforms mentioned above, this work proposes a thorough study of the mechanisms available in current state-of-the-art cloud-native tools and frameworks to enable isolation between slices of private 5G networks, with a specific focus on the \gls{5gc}. The central idea of the study is to explore mechanisms already embedded in edge/cloud platforms, such as the ability to allocate specific resources to certain \gls{vnf}s. This includes limiting resource usage, such as CPU time, network bandwidth, or memory, in addition to process prioritization.

A significant contribution of this work is the development of a methodology for experimentation that explores the use of said resource control mechanisms, individually and in combinations, providing guidelines to private 5G network operators on how to use them to ensure slice isolation. Additionally, the data obtained in this research has been made available and can be applied to the development of future data-driven applications for the management or control of 5G networks, such as those employing machine learning. One example is the work by \cite{Gupta:2019}, which uses simulated data to train machine learning models and could benefit from a dataset like the one made available in this study to obtain realistic models. The scripts developed for experimentation in this work are also publicly available and can be used to reproduce the research and/or create other experiments, such as comparing other applications with different 5G requirements and variations in slice configurations\footnote{https://github.com/maikovisky/open5gs}.

During the experiments, a hypothetical scenario was developed where a private network operates in a hospital complex with units in other cities. In this scenario, high-quality video calls occur between units, including doctors guiding a remote operation, making it necessary to prioritize the network of these calls over others. Resource limitations were applied, both individually and in combinations. For instance, we found that limiting CPU resources for non-priority slices benefits the priority slice, while memory limitation has no noticeable effects.

This work is organized as follows: In \autoref{sec2}, we discuss the operation of 5G networks. In \autoref{sec3} we introduce the environment where the experiments are conducted. The experiments themselves are detailed in \autoref{sec4}, where we analyze the key aspects of each. Finally, in \autoref{sec5}, we present the conclusions of this work and perspective for future work.

\section{5G Network Slicing and Isolation}\label{sec2}

The evolution of mobile networks has transformed aspects of everyday life, with each generation bringing advancements in coverage, mobility, security, and connectivity. The fifth generation, 5G, was developed with a focus on three main use cases: enhanced mobile broadband (\textit{eMBB}), ultra-reliable low-latency communication (\textit{URLLC}), and massive machine-type communication (\textit{mMTC}) \cite{SouzaNeto2021}. These innovations offer significant improvements in terms of speed, latency, and connection capacity. These advancements make 5G a central technology for supporting the growing demand for connected services, such as the Internet of Things (IoT), autonomous vehicles, augmented reality applications, and medical emergencies, among others. In addition to offering greater bandwidth, 5G introduces innovations that enable increased flexibility and efficiency in resource management, particularly through virtualization and network slicing \cite{Shorov2019}.

In this section, we review the main changes introduced by 5G networks, focusing on the network core, including the network functions for both the control plane and the user plane, as well as the isolation mechanisms and innovations that differentiate it from previous generations. We also present the main open-source 5G applications. The goal is to understand the operation of 5G, with an emphasis on network isolation, providing the foundation for building an adequate 5G network for experimentation. Finally, we present related work on the topic of 5G network slice isolation.

\subsection{Key Changes in 5G Networks}

When the \gls{3gpp} announced that Release 15 of the 5G standard would adopt \gls{sba}, it allowed the primary network functionality -- i.e., \gls{nf} -- to be designed as cloud-native components, leveraging advanced networking technologies such as \gls{nfv} and \gls{sdn} \cite{Scotece2023a}. Each of the 11 \gls{nf}s was modularized, enabling microservice implementations for each of these virtual functions \cite{Silveira2022}. This allows the orchestration of the 5G network by scaling specific functions according to network needs, as well as distributing certain functions closer to the edge, i.e., closer to the end user. The 5G standards also logically separates network functionality into \gls{cp} and \gls{up}, both within the \gls{5gc} and the \gls{ran}~\cite{Khaturia2021}.

The idea of adopting \gls{sba} is to enable the advantages of software development best practices for cloud computing environments in telecommunications, defining open standards for \gls{nf} integrations. This reduces costs, eliminates dependency on specific equipment or vendors, and opens the door for innovations by enabling improvements or individual studies for each component \cite{Luong2018}. In this work, we study the operation of slicing in 5G networks, but for this, it is necessary to understand the \gls{nf}s of both the \gls{cp} and the \gls{up}, which are discussed in this section, as well as open-source 5G implementations.

\subsection{Network Slice Isolation}

The concept of network slicing was introduced in the context of 5G networks to enable support for a wide variety of use cases with different requirements, utilizing a single shared infrastructure. In this scenario, multiple specialized network slices are instantiated to provide the required services through \gls{e2e} network partitions. These instances, also referred to as \gls{e2e} logical networks, are known as \gls{nsi}, as described in \cite{Gonzalez:2020}. According to the same reference, isolation is defined as: ``\textit{the property that services in one slice can operate without any direct or indirect influence from activities in other slices, and without unsolicited influence from infrastructure providers.}''

\gls{nsi}s must operate independently, as though they are separate networks, with no influence from one to another. When the same infrastructure is shared, this is called isolation, which is a fundamental property for 5G network slicing. A significant challenge is ensuring that failures in one \gls{nsi} do not propagate to others, preserving the guarantees of contracted services \cite{Gonzalez:2020}. It is also crucial to ensure that the requirements of each application running on a slice are individually observed so that the performance of one slice is not impacted by others.

\subsection{Control Plane}

\begin{figure}[H]

\centerline{\includegraphics[width=30em]{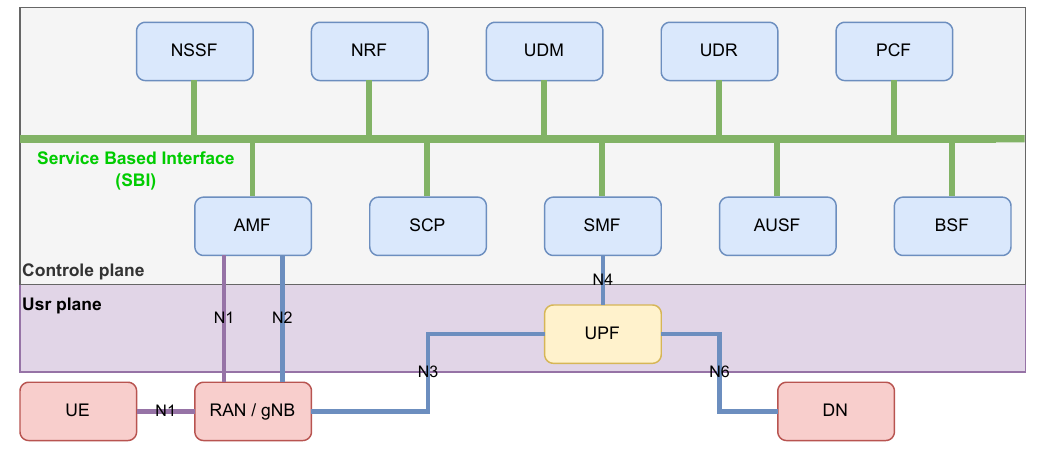}}
\caption{Service-based architecture of 5G core network.}
\label{fig:5gnfs}
\end{figure}

The control plane acts as the ``brain'' that regulates device operations based on the \gls{nf}s. In 5G networks, the control plane is responsible for dictating network behavior, relying on the policies and rules established for each \gls{nf}. This includes defining how a \gls{ue} operates within the network \cite{Ghaffar2023}. Figure \ref{fig:5gnfs} illustrates the division between the \gls{cp} and the \gls{up}, highlighting the positioning of the \gls{nf}s within this structure.

Each network function presented in Figure \ref{fig:5gnfs} must be registered, a task that falls under the responsibility of the \gls{nrf}. As a result, other network functions only need to know the \gls{nrf}’s address, as it is responsible for maintaining control over all services available in the network.

The \gls{amf} plays a fundamental role in 5G networks, being present in most signaling call flows. Its primary function is to mediate communication between the radio access network and devices, using the N1 and N2 interfaces, as shown in Figure \ref{fig:5gnfs}. The \gls{amf} is responsible for registering, authenticating, and managing the movement of devices between radio cells, as well as reactivating idle-mode devices. To ensure security, encrypted connections are employed \cite{Rommer2020a}.

The \gls{smf} manages sessions for end users (devices), which includes establishing, modifying, and releasing individual sessions, as well as allocating IP addresses for each session. Communication between the \gls{smf} and devices occurs indirectly through the \gls{amf}, which forwards session-related messages. Another role of the \gls{smf} is to configure traffic forwarding and apply traffic rules in the user \gls{upf} for each session \cite{Rommer2020a}.

The \gls{smf} establishes communication with the \gls{pcf}, which plays a critical role in two main areas: policy control for data sessions and policy control unrelated to data sessions. The latter allows service providers to manage user access to the network, enabling, for example, restrictions on geographical connection areas or defining permitted radio access technologies, through interaction with the \gls{amf} and the \gls{udr}. The \gls{udr} acts as the central database that stores network and user policies, providing data storage and access services for other network functions. Additionally, the \gls{pcf} communicates with the \gls{bsf}, which provides services for registering and deregistering data session information and informs other applications about which \gls{pcf} serves a specific session \cite{Rommer2020a}.

The \gls{udm} function is responsible for managing user subscription data and authenticating devices in the 5G network. Meanwhile, the \gls{ausf} authenticates devices using credentials from the \gls{udm} and generates cryptographic material to ensure secure updates. These functions play a crucial role in access management and security within the network \cite{Rommer2020a}.

The \gls{nssf} is responsible for selecting network slices based on a combination of single \gls{s-nssai} values, which are defined for the network, requested by the device, and permitted in the subscription. The \gls{s-nssai} identifies specific slices in the 5G network, and its subparameters, slice/service type (SST) and \gls{sd}, are used to distinguish these slices. The initial selection of the \gls{amf} is determined by the \gls{s-nssai} values requested by the device. The \gls{amf} can directly serve the device or use the \gls{nssf} to perform a new slice selection. A device can connect to multiple slices, each with its own specific functions, such as \gls{amf}, \gls{smf}, and \gls{upf}.

\subsection{User Plane}

The \gls{upf} ensures that specific devices remain accessible via the Internet, even while on the move. This function is responsible for processing data, generating traffic usage reports for the \gls{smf}, and performing packet inspection for policy decision-making. The \gls{upf} applies network policies, such as traffic redirection and data rate limitations. When a device is idle, it stores the traffic and reconnects it when necessary. Additionally, the \gls{upf} applies Quality of Service (\textit{QoS}) markings to prioritize packets in congestion scenarios \cite{Rommer2020a}. However, in the experiments conducted in this work, the open-source \gls{upf} implementations tested do not implement any type of prioritization; they simply forward packets in the order they are received.


The \gls{3gpp} defines a logical entity or node called \gls{gnb}, or \textit{Next Generation Node B}, which corresponds to the functionality of the radio network and is typically referred to as a radio base station when implemented as a deployable network product. The term \gls{gnb} is specifically used to refer to an NR base station connected to the 5G network. This node is designed to be scalable and flexible, meeting the growing demand for mobile broadband services \cite{Rommer2020a}.

As illustrated in Figure \ref{fig:5gnfs}, the \gls{gnb} is an integral part of the \gls{ran} in the 5G context. During the \textit{downlink}, the \gls{upf} is responsible for forwarding traffic destined for the \gls{ue}, encapsulating it in the GPRS Tunneling Protocol over the User Plane (GTP-U) through the N3 connection shown in Figure \ref{fig:5gnfs}. At this stage, the \gls{gnb} processes the received packets by unpacking them, encrypting the payload, and adding the PDCP (\textit{Packet Data Convergence Protocol}) and RLC (\textit{Radio Link Control}) headers before sending them to the user equipment. In the case of the \textit{uplink}, the \gls{gnb} identifies and separates control packets (RLC) from user data. The data is then decrypted and forwarded via the GTP-U tunnel to a \gls{upf} instance, completing the communication flow \cite{Singh:2023}.

\subsection{Open-Source 5G Implementations}

To develop and consolidate new concepts—such as network slicing—in the field of mobile networks, it is essential for researchers to establish a functional 5G network to evaluate the performance of new proposals and applications. Open-source implementations of the 5G core, such as Open5GS, Free5GC, and OpenAirInterface, emerge as an attractive and cost-effective option for validating applications. These implementations allow network functions to operate independently, facilitating orchestration in containers. By using cloud-native tools such as Kubernetes, it is possible to create a more realistic environment, distributing 5G network functions across multiple nodes and configuring scenarios with edge computing. Despite the continuous progress of open-source tools for 5G networks, challenges such as data traffic prioritization, connecting large numbers of \gls{ue}, and limitations in network slice selection, creation, and isolation still persist \cite{Lando2023}.

\subsubsection{Open5GS}

Open5GS is an open-source 5G core network that provides support for both NSA and SA operating modes. The 5G NSA core, in turn, is essentially an adaptation of a core for 4G networks, featuring important modifications such as the separation between the user plane and the control plane. The 5G SA core, on the other hand, is a fully functional 5G core compliant with release 17 of 3GPP \cite{Hakegard2024}.

As illustrated in Figure \ref{fig:5gnfs}, the components that comprise the control plane and the user plane are part of the 5G core integrated into Open5GS. These components represent the modular and flexible architecture of the solution, enabling its use in different deployment scenarios.

\subsubsection{Free5GC}

Free5GC is an open-source project developed for the 5G core network, receiving significant contributions from the National Chiao Tung University and evolving from the NextEPC project, which has since been discontinued. Due to the unavailability of commercial 5G devices in the market during its conception, Free5GC adopts the 4G protocol to communicate with UEs and the 4G base station. As a result, its authentication protocol is also based on 4G rather than 5G. Nevertheless, Free5GC implements 5G functionalities that comply with release 15 of 3GPP \cite{SouzaNeto2021}.

\subsubsection{OpenAirInterface (OAI)}

The OpenAirInterface project focuses on developing a comprehensive solution for 5G networks. It includes the radio access network (RAN) project, aimed at implementing a 5G radio access system, and the 5G core project, which provides a complete implementation of the control plane autonomously and in compliance with release 17 of 3GPP specifications. Additionally, the project incorporates an orchestrator, as well as monitoring and maintenance tools, for both the RAN and the 5G core \cite{OpenAirInterface2024a}.

\subsubsection{srsRAN}

srsRAN is a complete RAN solution developed according to 3GPP and O-RAN Alliance specifications. This platform offers a full L1/2/3 stack, characterized by portability and scalability, making it suitable for both embedded systems and cloudRAN-based environments. These features make srsRAN widely used in research and development in the field of mobile networks. Moreover, srsRAN supports integration with radio intelligent controller (RIC), PHY solutions, and other O-RAN-compliant hardware and applications, promoting flexibility and enabling the efficient use of third-party solutions. Although it provides a robust solution for establishing the connection between the UE and the core, core functionality is not within the scope of the srsRAN project \cite{srsRAN2024a}.

\subsection{Simulation Tools for UE and RAN}

In addition to the 5G implementations presented, there are tools that allow radio access networks to be simulated via software, as well as user equipment (UE). These tools include UERANSIM and my5G-RANTester.

\subsubsection{UERANSIM}

UERANSIM is a simulator for user equipment (UE) and radio access network (RAN) in 5G networks, designed as an open-source solution that simulates both 5G mobile devices and base stations. The project aims to enable testing and research on 5G networks. However, UERANSIM does not offer a complete implementation of the 5G physical layer, it have limitations in radio channel and physical layer modeling \cite{Rouili2024}. The 5G-NR radio interface is only partially implemented and simulated using the UDP protocol \cite{GUNGOR:2024}. The simulator includes the components nr-gnb and nr-ue, which represent, respectively, the gNodeB (gNB) and the UE. When establishing a connection with the 5G network, the UE creates a network interface responsible for tunneling with the user plane function (UPF).

\subsubsection{my5G-RANTester}

The my5G-RANTester is an emulation tool for the control and user planes of UE and gNB in 5G networks, focusing on the implementation of NGAP and NAS protocols according to 3GPP Release 15. This tool also allows researchers to study the functionalities of the 5G core and test its compliance with standards, and it is scalable for simulating many UEs. However, the wireless channel is not yet implemented. The my5G-RANTester leverages libraries and data structures from the Free5GC project \cite{Silveira2022}. Additionally, in \cite{Lando2023}, the tool was extended to include performance testing for the 5G core.

\subsection{Tool Comparisons}

A comparative study presented in \cite{Lando2023}, which evaluated three open-source 5G core implementations, showed that OpenAirInterface (OAI) was less mature, to the point that it hindered the execution of the proposed experiments. In contrast, Free5GC stood out for its better performance in user plane throughput, while Open5GS demonstrated greater stability in terms of the number and duration of UE connections, as well as efficient resource usage.

The connection of UEs to the 5G core uses wireless equipment, which proves to be costly for this work, highlighting the need for a gNodeB (Next Generation Node B) simulator. In \cite{Lando2023}, the my5G-RANTester was employed for testing purposes. However, this tool has limitations, such as the dependency between the UE and the gNodeB, which are initialized together, and the difficulty of dynamically increasing the number of UEs. Another disadvantage is the inability to separate the gNodeB and UEs into different nodes or dynamically create new UEs. Alternatively, UERANSIM, as suggested by the developers in \cite{Lee2024a}, simulates a radio connection and allows for the creation of a gNodeB independently of the UEs, facilitating the dynamic addition of new UEs \cite{GUNGOR:2024}.

\section{Related Work}

Several studies have explored 5G slicing testbeds and mechanisms to ensure performance and isolation guarantees. Esmaeily and Kralevska~\cite{Esmaeily:2021} provide a comprehensive comparison of 5G slicing environments, defining primary criteria such as support for SDN, NFV, and cloud computing, as well as dynamic orchestration and multi-tenancy. Secondary criteria include multi-RAT support, E2E slicing, ML integration, and open-source availability. Despite these advances, the authors observe that most evaluated testbeds offer only partial isolation or limit slicing to specific protocol layers.

Rouili et al.~\cite{Rouili2024} evaluate the fidelity of open-source 5G standalone testbeds by comparing simulation/emulation tools with SDR-based deployments. Although their focus is on RAN performance rather than core-level slicing, they underscore the challenges of achieving realistic 5G environments using open-source components—challenges that similarly affect core slicing deployments. Similarly, Garcia-Aviles et al.~\cite{GarciaAviles2018a,GarciaAviles2020b} propose the POSENS protocol, which enables E2E slicing through modifications to PDCP and RRC layers and traffic redirection at a load balancer node to enforce bandwidth limitations between slices. In contrast, Shorov et al.~\cite{Shorov2019} and Dzogovic et al.~\cite{Dzogovic2018} present early testbeds focused on slice selection and deployment across virtualized cores, though without addressing slice isolation explicitly.

More recent work investigates slicing from a performance assurance perspective. Nikolaidis et al.~\cite{Nikolaidis2023} address per-user QoS guarantees in RAN slices using a multiclass queuing model for bandwidth provisioning. While their focus is on intra-slice differentiation and SLA negotiation, our work targets inter-slice performance isolation using infrastructure-level controls. De Simone et al.~\cite{DeSimone2024} present a formal approach to optimizing performance and availability in resilient 5G architectures through stochastic modeling and empirical validation using Open5GS. Unlike their SLA-driven redundancy planning, we concentrate on mitigating slice interference through direct resource controls in edge/cloud-native environments.

Other studies adopt broader system-level perspectives. Minardi et al.~\cite{Minardi2023} explore SDN-based control strategies for integrated terrestrial and non-terrestrial networks, focusing on reactive and proactive traffic rerouting rather than resource isolation. Adeppady et al.~\cite{Adeppady2023} propose the iPlace heuristic to reduce microservice interference and deployment time in constrained cloud environments. Their work highlights the importance of resource-aware orchestration, aligning with our focus on runtime performance stability in private 5G infrastructures. Muro et al.~\cite{Muro2023} investigate the impact of the Noisy Neighbour phenomenon in virtualized 5G cores and apply machine learning for detection and forecasting. While they propose reactive mitigation, our work emphasizes proactive enforcement of CPU and bandwidth limits to avoid contention altogether.

At the orchestration layer, Ait Aba et al.~\cite{AitAba2024} enhance Kubernetes deployment strategies by integrating Genetic Algorithms to solve the Virtual Network Embedding problem and reduce failed slice deployments. Whereas their contribution targets slice admission efficiency, our approach deals with maintaining performance guarantees post-deployment, reinforcing the need to adapt general-purpose orchestrators to telecom-specific requirements.

Taken together, these studies demonstrate the growing interest in network slicing and resource management for 5G. However, they also reveal gaps in effective, reproducible solutions for slice isolation in open-source, resource-constrained environments. Our work contributes to this space by experimentally evaluating native resource control mechanisms—such as CPU quotas, bandwidth shaping, and prioritization—in Kubernetes-managed private 5G deployments, offering a practical and reproducible path toward performance-aware slice management.

\section{Experimental Environment Configuration}\label{sec3}

The 5G technology introduces the possibility of creating private networks, allowing the development of smaller-scale infrastructures, particularly advantageous for testing and experimentation environments. In this context, open-source tools stand out as a promising choice, as they offer the freedom to modify the code, enabling the validation of concepts that would not be feasible in proprietary networks. During this study, when evaluating open-source solutions, a certain immaturity in these tools was identified \cite{Silveira2022, Lando2023}.
One of the main innovations of 5G, network isolation, exemplifies this immaturity, as discussed throughout this section. Nevertheless, this functionality also paves the way for the implementation of mechanisms that enable the isolation of network slices, the central goal of this work. This section addresses the open-source projects available for the 5G core, presents auxiliary tools necessary for the deployment of a private 5G network in a edge/cloud environment, and discusses mechanisms available to achieve slice isolation. Finally, the structure used to conduct the experiments is described.

\subsection{Orchestration}

Microservices have emerged as an evolution of traditional monolithic applications, which operated as a single process on a robust computer, requiring hardware upgrades to scale. With the advent of microservices, applications are divided into multiple specialized processes, each with its own development cycle, dependencies, and library versions. This independence, while advantageous, can lead to incompatibilities when processes are executed on the same operating system \cite{Luksa2020}.

The concept of containers was developed to isolate microservices, allowing each to have its own dedicated environment, ensuring application isolation from one another. Containers leverage fundamental capabilities of the Linux kernel, particularly namespaces, which create separate views of process identifiers, users, file systems, and network interfaces \cite{Poulton2019}. As a result, applications can be managed individually and no longer need to share the same computer, enabling scalability and requiring configuration of communication among them. As the number of microservices increases, so does the management complexity, making the automation of this process essential \cite{Luksa2020}.

An orchestrator is a system that deploys and manages applications based on microservices, dynamically responding to changes. Kubernetes is an open-source application orchestrator created by Google, based on years of experience running containers at scale. This project is part of the CNCF, which has helped consolidate its role as the industry standard for the deployment and management API for cloud-native applications. With the ability to operate on any cloud or on-premises datacenter, Kubernetes abstracts the infrastructure, simplifying the creation of hybrid clouds and facilitating migration between different cloud service providers \cite{Poulton2019}.

Kubernetes is a tool capable of managing a set of devices (cluster) composed of a primary node, called the control plane, and a set of machines or worker nodes. These nodes can include high-capacity physical servers, virtual machines, cloud instances, or edge devices with resource constraints, such as the Raspberry Pi. In the control plane, all orchestration functionalities that require some form of ``intelligence'' are implemented, including scheduling, automated scaling, and continuous updates of the applications running on the worker nodes \cite{Poulton2019}. Currently, Kubernetes is the de facto standard for microservice orchestration and is the system chosen to manage the experiments conducted in this research.

\subsection{Auxiliary Components}

In addition to the basic components for nodes or the control plane, auxiliary components were used to improve communication between pods, capture and visualize metrics, and run applications.

\subsubsection{Container Network Interface}

Kubernetes \gls{cni} plugins are essential for the operation of containerized environments, significantly impacting network performance, scalability, and resource management. The choice of \gls{cni} directly affects network efficiency in Kubernetes, influencing aspects such as bandwidth and latency. Plugins like Antrea, Flannel, Cilium, and Calico each have their own advantages and disadvantages \cite{Dakic2024a}.

Antrea is a \gls{cni} plugin designed for Kubernetes that leverages Open vSwitch (OVS) to enhance network performance and provide robust security. It offers features such as network policies, traffic engineering, and observability. According to \cite{Dakic2024a}, the use of hardware offloading contributes to optimized performance, making it suitable for applications requiring high throughput and low latency.

Flannel is a simple and easy-to-configure CNI that uses VXLAN encapsulation for overlay networks, facilitating communication between pods on different nodes. According to \cite{Dakic2024a}, it does not offer the highest throughput among CNIs and has limitations in advanced functionalities like network policies and traffic segmentation. Due to its simplicity, it is ideal for small and medium-sized clusters.

Cilium is a CNI plugin for Kubernetes that uses eBPF to optimize network security and performance, enabling effective management of complex policies and detailed traffic observation, making it suitable for high-performance applications \cite{Dakic2024a}. The use of eBPF allows the Linux kernel to become programmable, enabling the execution of code snippets quickly and safely within the operating system \cite{Duong2023a}.

According to \cite{Dakic2024a}, Calico is efficient, standing out for its ability to enforce strict network policies and its scalability. It operates at Kubernetes layer 3 and offers both encapsulated and non-encapsulated networking options. Calico's architecture allows for the management of a large number of network policies without compromising performance, making it ideal for environments requiring high performance and security. Its integration with BGP also enables the use of advanced routing features, especially in complex networks.

In the performance evaluation by \cite{Dakic2024a} of the CNIs mentioned above, they show similar results. Regarding UDP latency, Flannel stands out. In terms of bandwidth, the CNIs deliver excellent performance for packets up to 1472 MTU but perform worse with larger packets. In terms of kernel optimization, Calico shows significant improvement compared to the others in scenarios with 2.5 Gbit/s. In this work, we opted to use Cilium due to its more efficient bandwidth limiter, as described in \autoref{chap3:limitband}, and also because of its use of eBPF, which may be utilized in future work.

\subsubsection{Observability and Monitoring}

Prometheus is an open-source monitoring and alerting platform, developed in 2012 and released by SoundCloud in 2016. It collects time-series data from data sources at defined intervals, storing it and identifying it by metric name and key/value pairs. Prometheus uses an HTTP-based pull model to collect real-time metrics and employs PromQL for flexible queries and real-time alerts. Unlike black-box monitoring, Prometheus promotes white-box monitoring, providing detailed insights into the state of microservices. Its main components include the Prometheus server, client libraries, Alert-manager, and exporters. Grafana is also very often used for displaying Prometheus metrics into dashboards \cite{Sukhija2019a}.

As the only system directly supported by Kubernetes and also the standard in cloud-native ecosystems\footnote{Prometheus FAQ \url{https://prometheus.io/docs/introduction/faq/}}, Prometheus was chosen as the monitoring tool and played an essential role in capturing data during the experiments. With it, it was possible to visualize information about the memory, CPU, and network traffic behavior of each pod, as well as to generate comparative charts for network and CPU usage, presented in section \autoref{sec4}.

\subsubsection{Infrastructure as Code}

Infrastructure as Code (IAC) allows defining and managing infrastructure through code, treating operations as software, including hardware elements \cite{Brikman:2022}. In Kubernetes, applications are defined by objects represented in the API. These objects include types for deployment, instance execution, and service exposure. Definitions are made in YAML or JSON files, formats that allow object serialization. YAML is human-readable, while JSON is suitable for data exchange between applications. Deploying an application may involve creating manifests that specify these definitions and associated services \cite{Luksa2020}.

Terraform is an open-source tool developed by HashiCorp. It enables the use of a simple declarative language to manage infrastructure in public and private clouds with just a few commands, simplifying tasks previously performed manually \cite{Brikman:2022}. This tool was used to create the environment needed to run the experiments, in which manifests were specified to deploy each function of the Open5GS core and their respective configurations. This approach makes it possible to reproduce the experiments in other environments.

\subsection{Mechanisms to Achieve Slice Isolation}

To achieve the objectives of this work, three mechanisms for controlling computational resources were selected: (\textit{i}) resource reservation (CPU and memory) in Kubernetes, (\textit{ii}) bandwidth limitation offered by Kubernetes in conjunction with the Cilium CNI, and (\textit{iii}) CPU prioritization using the nice command in Linux. These mechanisms were used, individually and in combination, in the experiments to prioritize 5G network slices and ensure some level of isolation. The operation of each of these mechanisms is detailed below.

\subsubsection{Resource Reservation in Kubernetes}

In Kubernetes, it is possible to define the amount of resources required for a container, such as CPU and memory (RAM). By specifying these resources, the \textit{kube-scheduler} uses the information to decide on which node the pod will be allocated. When a limit value is set, the \textit{kubelet} enforces these limits to ensure that the running container does not exceed them, which could cause errors, such as out-of-memory conditions, or even terminate the pod. Additionally, it is possible to reserve a minimum amount of resources, ensuring the pod has sufficient resources for its execution.

CPU resource reservation can be set to 1 CPU, which is equivalent to a physical CPU or 1 virtual core, represented as 1.0 or 1000m (millicpu), indicating a full CPU. To reserve a fraction of the CPU, values such as 0.1 or 100m can be used, corresponding to 10\% of a CPU.
For memory, resource allocation is specified in bytes, using integer values. Suffixes like M and G can be used to indicate megabytes and gigabytes, respectively. In this work, we use Kubernetes' API to simplify the establishment of resource limits and requests for the 5G components, but the actual enforcement of these constraints is implemented within the underlying container management system with \textit{cgroups}. \textit{Cgroups} are a generic Linux kernel capability that manage resources like CPU, memory, and storage, and could be used regardless of Kubernetes.

\subsubsection{Bandwidth Limitation}
\label{chap3:limitband}

To enforce bandwidth limits on pods, Kubernetes provides support for traffic shaping through an experimental CNI plugin for bandwidth control. In this model, the operator applies an annotation to the pod to limit inbound and outbound traffic. In \cite{Vibert2022}, it is stated that limiting network consumption is a challenge, and the use of the CNI plugin has limitations. Inbound traffic is managed using packet queuing in \textit{qdisc}, which is processed by the Token Bucket Filter (TBF) algorithm before entering the pod. This algorithm uses tokens that correspond to the amount of traffic that can be sent at once when the bucket is full. When no tokens are available, packets are queued up to a limit. As a result, TBF may hold packets until there are enough tokens, increasing latency.

For outbound traffic from the pod, \cite{Vibert2022} states that the CNI plugin redirects traffic to an Intermediate Functional Block (IFB) to apply traffic shaping. Consequently, an additional artificial hop is introduced. This experimental plugin increases resource consumption on the nodes and reduces performance gains in networks with modern NICs, as TBF is not suitable for multi-core or multi-queue scenarios. Additionally, the use of TBF adds hops and latency to the traffic.

According to \cite{Vibert2022}, Van Jacobson proposed a new approach to network congestion management during a conference, replacing queue-based models with a mechanism called Timing Wheels. This model operates by assigning each packet a timestamp called Earliest Departure Time (EDT), which determines the earliest time the packet can be sent based on transmission policies and the specified rate. This method not only regulates network flow but also reduces unnecessary delays and avoids packet loss. Compared to traditional models, Timing Wheels offer significant advantages in network traffic performance while minimizing CPU usage on the nodes.

This approach is used by Cilium for bandwidth management, employing the Earliest Departure Time (EDT) logic to improve efficiency in flow and congestion control. This implementation eliminated the need for IFB, reducing the latency associated with the implementation of \gls{cni} plugins. Additionally, Cilium leverages multi-core and multi-queue capabilities, ensuring that rate-limiting does not compromise the overall system performance. The latency test comparison between EDT and TBF presented by \cite{Vibert2022} showed a reduction of up to 4 times in latency for EDT. The more robust development of traffic control and the greater efficiency of this model were factors that contributed to the choice of Cilium as the CNI for the Kubernetes cluster. This is the mechanism used in this work to limit the bandwidth of 5G components in order to achieve slice isolation.

\subsubsection{CPU Prioritization (via nice)}

Linux has a scheduler that allows the allocation of CPU fractions to tasks, divided into two priority queues: one for active tasks and another for expired tasks. Active tasks are organized by priority order, and at the end of their execution, they are moved to the expired tasks queue. When the active tasks queue is emptied, the tasks from the expired queue are transferred back to the active queue \cite{Silberschatz2013}.

The nice value determines the time slice, that is, the time a task will occupy the processor. According to \cite{Silberschatz2013}, nice values range from -20 to 19, corresponding to time fractions between 800 ms and 5 ms, with a value of zero equivalent to 100 ms. Thus, the percentage of CPU usage will follow the formula presented in Equation \ref{eq:timeslice}:

\begin{equation}
fraction(i) = \frac{timeslice(i)}{ \sum_{\forall j}^{}timeslice(j)}
\label{eq:timeslice}
\end{equation}

Where \textit{timeslice(i)} is the amount of time, in milliseconds, allocated for task (i). In turn, the sum ($\sum_{\forall j}^{}timeslice(j)$) represents the total amount of time allocated for all tasks.

\subsection{Experiment Environment}

To investigate the proposed isolation mechanisms, an experimental environment was established simulating a 5G network in a Kubernetes cluster available at our university\footnote{Details on the Kubernetes cluster of UFRGS are available at https://computernetworks-ufrgs.github.io/k8s-cluster/}, monitored by Prometheus to extract metrics from each experiment and visualize them through Grafana. This environment also aims to bring the user plane, or UPF, closer to the end client, simulating an edge scenario. In contrast, the remaining functions of the 5G core are centralized, as if they were located in the cloud.

\begin{figure}[H]
    \centerline{\includegraphics[width=1\linewidth]{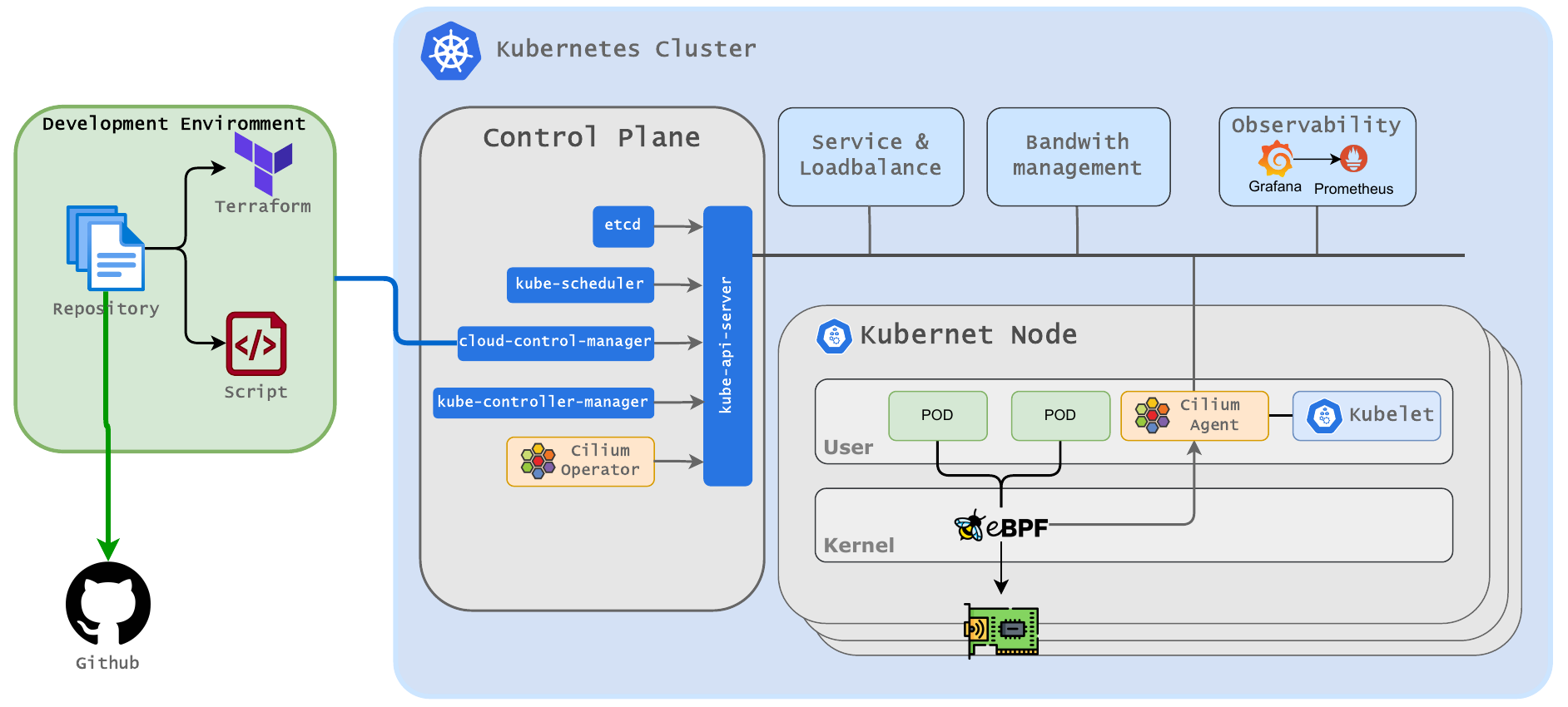}}
    
    \caption{Experiment Environment Simplified Architecture}
    \label{fig:topologiakubernet}
\end{figure}

Figure \ref{fig:topologiakubernet} represents the basic architecture of the Kubernetes cluster, detailing both the control plane node main components and those of the worker nodes. Additionally, the elements of Cilium, which manages the cluster's network, are incorporated, as well as the operators responsible for management and the agents that control the network interface of each node. It is noteworthy that communication with the network interface of each node is carried out using eBPF. Figure \ref{fig:topologiakubernet} also describes the development environment, responsible for applying Open5GS configurations via Terraform and executing the necessary scripts to initiate the experiments and collect the results. This environment operates externally to the cluster and communicates with it through the Kubernetes API.

\begin{figure}[h]
    \centerline{\includegraphics[width=1\linewidth]{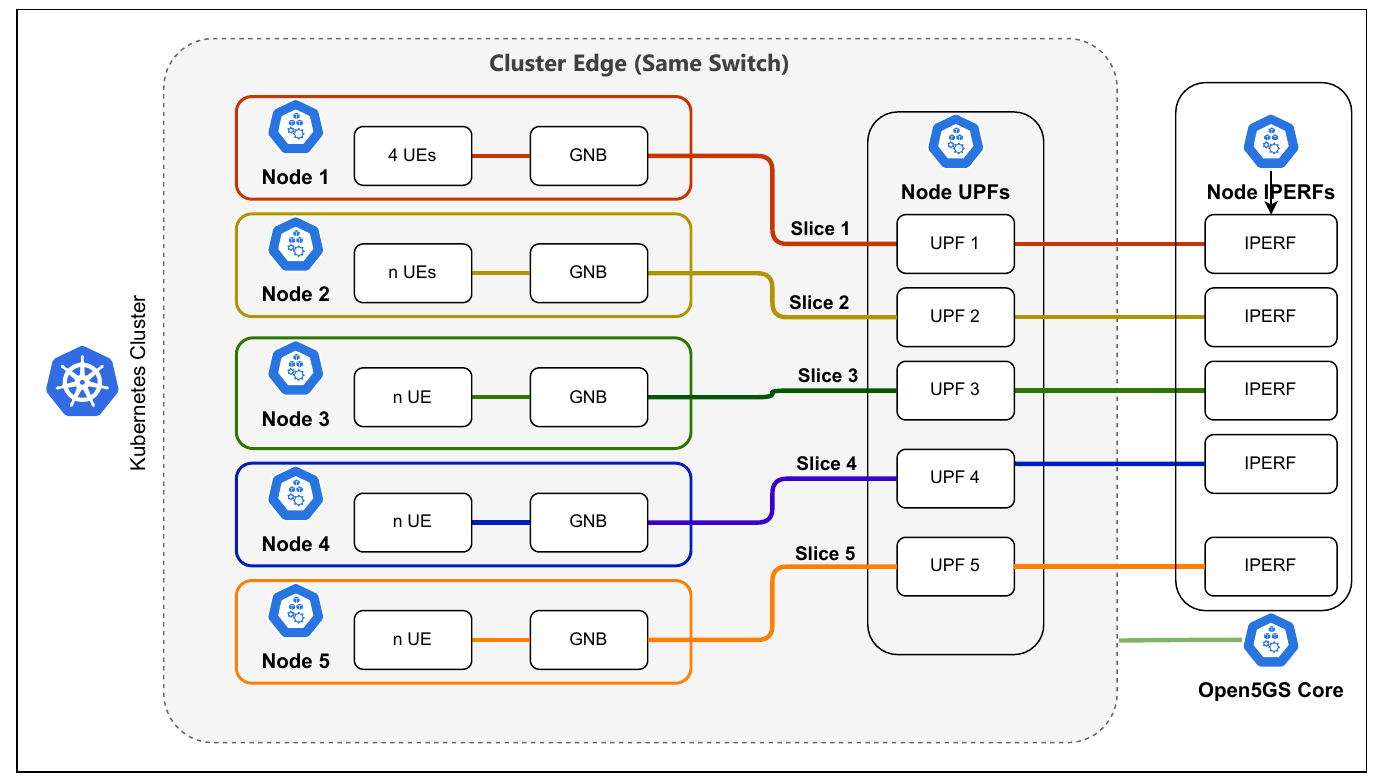}}
    \caption{Experimentation Environment Topology}
    \label{fig:topologia}
\end{figure}

Figure \ref{fig:topologia} shows the topology of the experimental environment, which consists of five edge nodes (i.e., resource-constrained\footnote{The hardware profile of each edge node is characterized as follows: a mini computer -- small chassis of 12 cm (width) by 4 cm (height) by 12 cm (depth) -- equipped with an Intel Core i7-5500U CPU @ 2.40GHz processor, 4GB of RAM DDR3 @ 1600MHz, 128GB SSD hard drive, and an onboard Realtek RTL8111 Gigabit Ethernet network card.} physical machines) dedicated to simulating the UEs and the gNB. All edge nodes host UEs/gNBs that generate traffic in specific network slices and are connected to the same physical switch, simulating a low-latency environment similar to an actual edge network. The edge node that hosts the UPFs network function, responsible for routing traffic from UEs over the 5G network to the Internet respecting slice configurations, is also connected to the same switch and has the same hardware profile.

By concentrating all UPFs on a single edge node, it is expected that resource contention occurs between the UPFs during the experiments, allowing the observation of the impact of this contention or interference among different slices. At the same time, the presence of this machine connected to the same switch brings the UPF closer to the edge of the cloud, which characterizes the concept of edge computing. Meanwhile, the remaining network functions of the \gls{5gc} are deployed on server nodes external to the switch, simulating a higher-latency environment such as cloud.

To simulate the UEs and the gNB, the UERANSIM was initially used, which allows the simulation of a single gNB with multiple UEs connected. However, during the experiments, due to implementation issues, one or more UEs occasionally lost connection and failed to reconnect, invalidating the experiments. Ultimately, the my5G-RANTester was chosen, which proved to be more stable, although it required one gNB for each UE.

\begin{figure}[H]
    
    \centerline{\includegraphics[width=\linewidth]{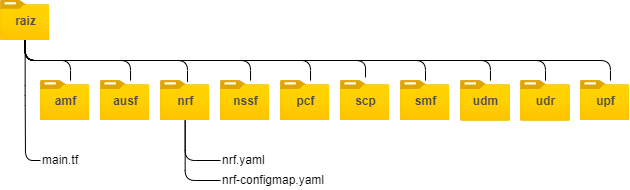}}
    \caption{Path Organization for Terraform}
    \label{fig:organization}
\end{figure}

To deploy the 5G core functions using Open5GS, \gls{iac} was employed through Terraform, enabling the simplified reproduction of the test environment. All codes, manifests, and scripts are available in a public GitHub repository\footnote{https://github.com/maikovisky/open5gs}. Figure \ref{fig:organization} illustrates the organization of the Open5GS configuration, where the main.tf file contains a set of Terraform commands that, in practice, access each folder and apply the manifests to Kubernetes. These manifests configure the network services, applications, and monitoring of each VNF.

To capture the metrics, Prometheus was used in conjunction with Grafana, which is the monitoring tool for the Kubernetes cluster. After the experiment, the data is extracted from Prometheus and stored in a non-relational database outside the cluster. This procedure is used to unify the results of the same experiment and calculate more detailed statistics (averages, standard deviation, etc.) from the metrics.

A script was developed to standardize each experimental run. This script restarts the 5G core and then configures the UPFs according to the mechanisms chosen for the specific experiment. It initiates the pods, which contain three containers: the first obtains a number from a specific slice and executes the my5G-RANTester command to connect to the corresponding slice; the second container runs the iPerf command, a tool used to generate traffic and measure bandwidth; and the third container uses Telegraf, an agent that assists in collecting latency metrics between the UE and the iPerf server, saving the metrics in Prometheus. The script runs the experiment for 20 minutes, terminating all pods with UEs at the end of the period. Finally, the script stores the Grafana charts on the local disk and saves the Prometheus data into a non-relational database. This procedure ensures that experimental data is not lost in case Prometheus needs to delete older data.

\begin{figure}[h]
    
    \centerline{\includegraphics[width=0.7\linewidth]{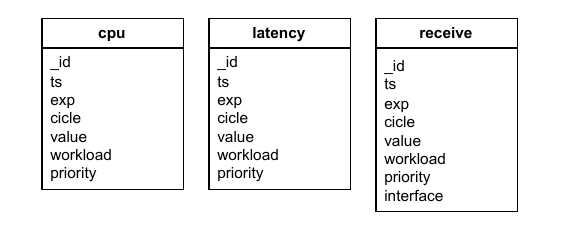}}
    \caption{Structure of the Collections Where CPU Metrics Are Stored}
    \label{fig:collections}
\end{figure}

Each measurement is stored as a record in a collection corresponding to the specific metric (CPU, latency, and received bytes). Figure \ref{fig:collections} represents the structure of these collections. This record contains the experiment number, the run or cycle, the metric value, the interface or workload, the time, which ranges from zero to 1200 seconds (or 20 minutes), and the priority, indicating whether it is the priority slice or others. This structure allows grouping of low priority slices when the data is analyzed.

A script was also developed to generate statistical data from the information stored in the database and create corresponding graphs. New collections are created to contain overall averages for each experiment, averages for the last 5 minutes (when the experiments reach the peak of traffic), and averages grouped by time. In these collections, the mean, minimum value, maximum value, standard deviation, and data count are calculated. Additionally, the workload, experiment, time, and whether the slice is primary or secondary are stored.
The configurations and scripts were saved in our open GitHub repository, which can be used by the community to reproduce the experiments or create new ones.

\subsection{Lessons Learned}

Setting up the environment required considerable time, during which various unforeseen events occurred, such as power outages, machine overheating, bugs in the tools used, and the need for configuration adjustments. The use of pods isolates the application under Kubernetes control, which complicates execution and adjustments to the applications, as it is necessary to wait for the pod to restart with the new configuration. To work around this issue, a sleep function was added to the command field of the pod manifest. This approach prevents the container from being immediately terminated when an error occurs or when it stops running. Subsequently, a Kubernetes command is used to access the pod's command line, allowing manual correction of configuration issues.

Initially, the UERANSIM tool was selected to simulate the UEs, as it enables the simulation of an RAN and the connection of multiple UEs. During the experiments, a reduction in network traffic was observed. Since resource contention was high, it was initially assumed that this reduction was due to the impact of resource contention. However, even in situations where there was no contention, the reduction in network traffic persisted.

Upon investigating the resources, it was found that UERANSIM had a bug that disconnected UEs from the 5G network and prevented them from reconnecting, even when trying to force a new connection. After multiple attempts to mitigate this issue, the decision was made to use the my5G-RANTester, which proved to be more stable. However, this tool does not allow the simulation of a RAN for multiple UEs, which is a limitation compared to UERANSIM.

Using pods to simulate a UE and reusing the same manifest presents the challenge of assigning a unique IMSI number to each UE. These numbers need to be exclusive and must meet the specific requirements of each network slice in which the UE is inserted. As a solution, a Python script was developed that uses the pod's IP address as a base to generate a unique IMSI number. This number is then registered in the database queried by the 5G core, and the application's configuration files are modified to use the IMSI assigned to the respective pod.

\FloatBarrier
\section{Experiments}\label{sec4}

The results of the experiments, conducted with the infrastructure described earlier, are presented in this section. CPU usage, memory usage, inbound and outbound network traffic, and latency for each network slice were measured. Each experiment involved modifying one or more mechanisms, analyzing the behavior of each slice to identify the optimal configurations for prioritizing a specific slice. Initially, the methodology designed for the experiments is described. Then, the results of the baseline experiments, without modifications to the mechanisms, are presented. Subsequently, individual changes to CPU limits, CPU prioritization, and bandwidth limitation are introduced, followed by combinations of these properties.

\subsection{Experimentation Methodology}

The objective of the experiments is to evaluate the ability to prioritize network slices at the edge by adjusting parameters available in a Kubernetes cluster, such as CPU allocation, bandwidth limitation, and, in the Linux operating system, processing prioritization using the \textit{nice} command. Parameter adjustments are applied exclusively to the UPF network function, which is responsible for the user plane. Therefore, we have designed a methodology that allows us to understand the impact of resource control mechanisms in slice isolation in the \gls{5gc}, as follows: (\textit{i}) first we conduct a set of baseline experiments with no resource restriction mechanisms in place to demonstrate and measure interference among slices, (\textit{ii}) then we apply resource restrictions one-by-one in varied configurations and measure their effectiveness to isolate slices, and (\textit{iii}) finally we apply combinations of the restrictions applied before to understand the combined effects of these measures.

For the experiments, pods are created to simulate UEs, mimicking the behavior of a camera transmitting high-resolution video for a remotely assisted medical operation. Each pod consists of four containers, with the initialization container always executed first, as per Kubernetes’ standard operation. This initialization container is responsible for configuring the corresponding network slice and defining the UE number in the 5G network. Only after its execution is complete do the other containers in the pod start simultaneously to perform the main simulation operations.

Among the containers executed after initialization, the first is responsible for simulating the UE, including running the base station (gNodeB). This container establishes a connection with the 5G core in the corresponding network slice using a GTP tunnel that connects the UE to the UPF. The second container is tasked with simulating video transmission, using the IPerf command to reproduce the traffic generated by a camera. The third container performs network latency testing. To do this, it runs Telegraf, which relies on the ping command to measure network latency. This latency measures the round-trip time (RTT) of the network. This modular structure allows each pod to efficiently emulate the behavior of devices connected to a 5G network, encompassing both data transmission and network performance monitoring.

During the experiments, several UEs (or pods) are executed over a 20-minute period. Initially, to represent the high-priority slice (Slice 1), 4 pods are started simulating 4 cameras streaming high-quality video at an operating room. Video is streamed over the edge-located \gls{upf} and delivered to a simulated cloud service (iPerf server). UEs connected to lower-priority slices (Slices 2 to 5) serve as background traffic and generate the same traffic patterns as the priority ones. In Slices 2 to 5, at every 4 minutes of experimentation, a new UE is added, while the pods are allocated to the node corresponding to the slice, as illustrated in Figure \ref{fig:topologia}. Video transmission is simulated using the isochronous\footnote{https://manpages.debian.org/bullseye/iperf/iperf.1.en.html} option of IPerf, configured with 60 frames per second at a bitrate of 45 Mbits per second, and no variation in frame size.

At the end of an experiment, each low-priority network slice has 5 UEs in operation, totaling 225 Mbps per slice. Slice 1, configured as the priority slice, maintains a constant rate of 180 Mbps throughout the experiment, receiving data from 4 UEs. However, at the end of the 20-minute period, the total sum of transmission rates across all clients reaches 1080 Mbps, exceeding the network link's limit (1 Gbps). This scenario forces resource contention, highlighting the need for efficient traffic management.

A total of 26 experiments were conducted, with each experiment executed 10 times. Metrics were collected by Prometheus every 30 seconds over the 20-minute duration of each experiment, resulting in 41 samples per slice and per experiment, totaling 2050 samples per experiment for each metric analyzed. The captured data includes the volume of received data, latency, CPU usage, and memory usage. In the initial experiments, while configuring the structure and parameters, it was found that imposing memory limits had no significant impact. All UPF network functions maintained constant levels of memory usage. Consequently, memory limitation was excluded from subsequent experiments, avoiding additional experimental runs deemed unnecessary.

\FloatBarrier
\subsection{Baseline Experiments}\label{sec:exp-baseline}

\begin{figure}[!htbp]
\centering
\includegraphics[width=\linewidth]{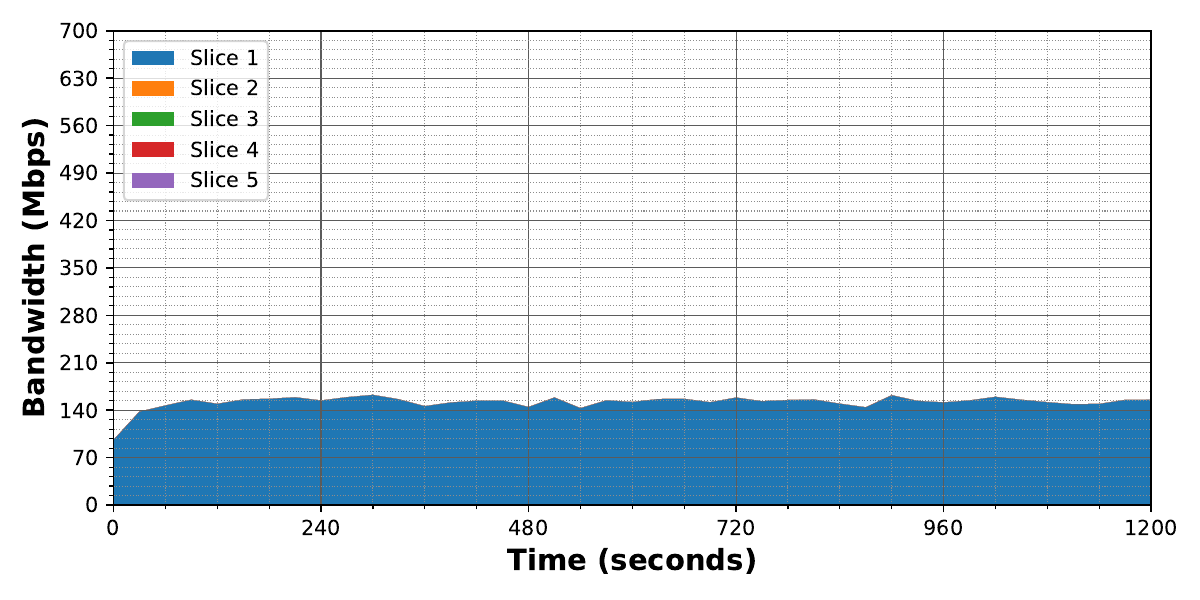}
\captionof{figure}{Network traffic behavior in Experiment 1}
\label{fig:exp01}
\end{figure}

The first five experiments investigated the behavior of each network slice in a scenario without restrictions for the UPFs, meaning no specific strategy was applied to ensure isolation between slices. In each experiment, a new slice was added, totaling five UPFs. The first experiment, illustrated in Figure \ref{fig:exp01}, was conducted with only one slice. It was observed that Slice 1 achieved an average data rate of 158.53 Mbps during the 20-minute (1200 seconds) experiment, establishing the ideal average rate for the priority slice.

\begin{figure}[!htbp]

\centering
\centering
\includegraphics[width=\linewidth]{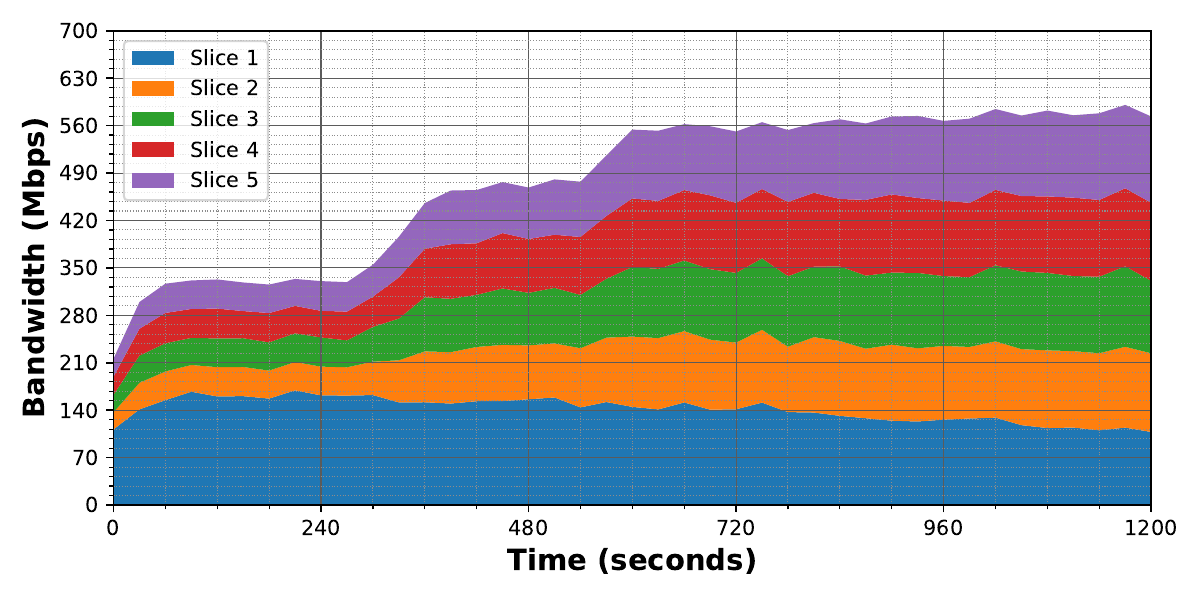}
\captionof{figure}{Network traffic behavior in Experiment 5}
\label{fig:line-receive-baseline-exp5}
\end{figure}

Conversely, Figure \ref{fig:line-receive-baseline-exp5} illustrates the degradation of the priority slice (Slice 1) over time when all network slices are in use. The increase in the number of UEs every 4 minutes in the other slices causes the system to reduce the traffic received by the iPerf server starting at 720 seconds.

\begin{figure}[!htbp]

\centering
\includegraphics[width=\linewidth]{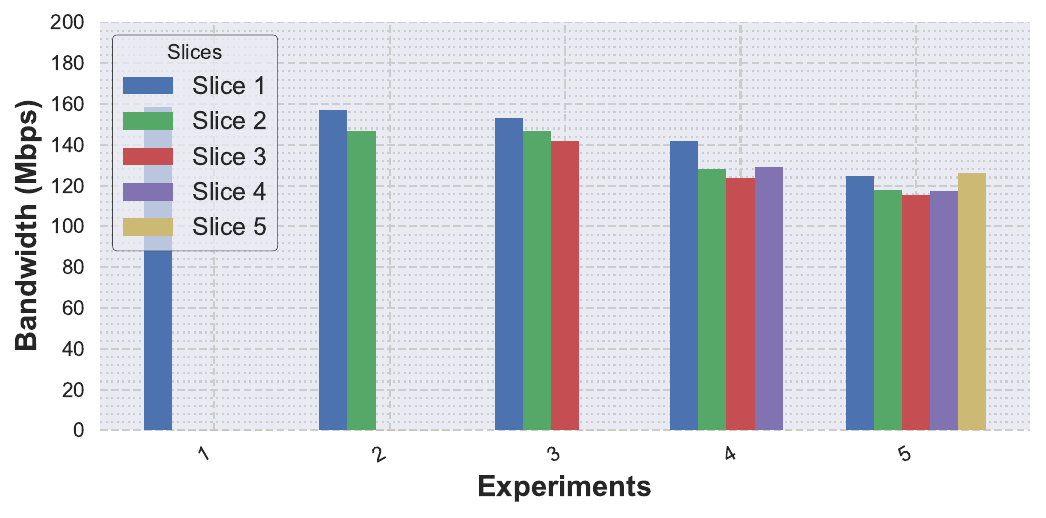}
\captionof{figure}{Network traffic in UPF for Experiments 1 to 5 - Last 5 minutes}
\label{fig:plot-baseline}
\end{figure}

Figure \ref{fig:plot-baseline} presents a comparison between the experiments during the last five minutes, which represents the worst situation in terms of interference among slices, illustrating the impact of adding each new slice. By analyzing these graphs, it can be seen that the priority slice maintains the average bitrate of approximately 158 Mbps when it is the only one generating traffic. In Experiments 2 and 3, a slight degradation in bandwidth is observed, with values of 156 Mbps and 153 Mbps, respectively. The introduction of the fourth slice results in a more pronounced degradation, reducing Slice 1's rate to about 141 Mbps. When the fifth slice is added, the rate drops to 124 Mbps, a reduction of more than 20\% compared to the ideal scenario (Experiment 1).

\begin{figure}[!htbp]

\centering
\includegraphics[width=0.9\linewidth]{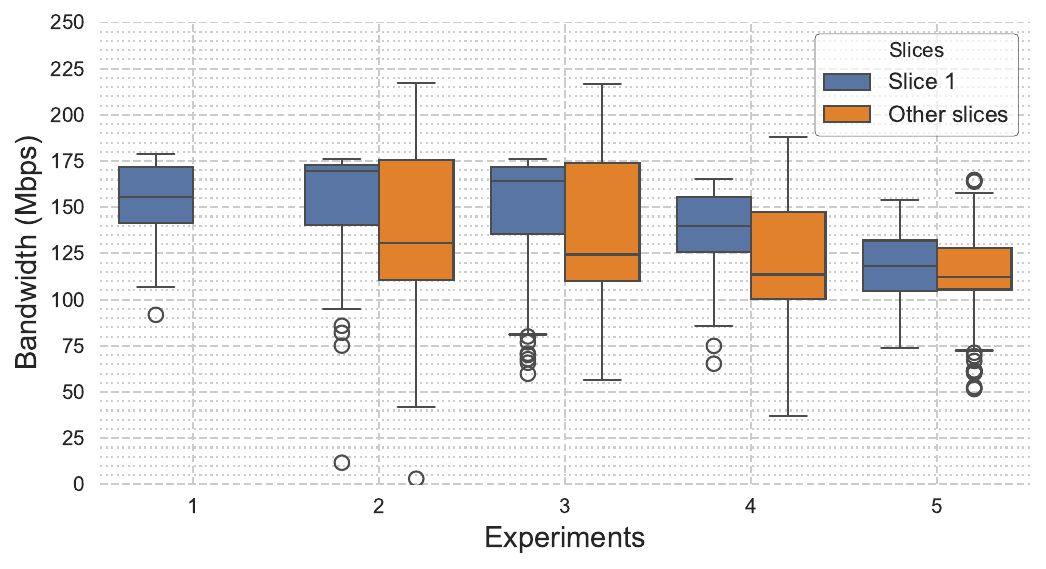}
\resizebox{\columnwidth}{!}
{\begin{tabular}{c|cccc|cccc}
\toprule
\multirow{3}{*}{Experiment} & \multicolumn{4}{c}{\textcolor{slice1}{\rule{1em}{1em}} Slice 1} & \multicolumn{4}{|c}{\textcolor{others}{\rule{1em}{1em}} Other slices} \\
& Mean & Median & Std. Dev. & IQR & Mean & Median & Std. Dev. & IQR\\
& (Mbps) & (Mbps) & (Mbps) & (Mbps) & (Mbps) & (Mbps) & (Mbps) & (Mbps)\\
\midrule
1 & 154.04 & 155.42 & 19.91 & 30.20 &  &  &  &  \\
2 & 151.99 & 169.50 & 29.18 & 32.33 & 139.15 & 130.68 & 48.13 & 64.84 \\
3 & 149.77 & 164.18 & 31.04 & 36.45 & 140.36 & 124.45 & 40.86 & 63.71 \\
4 & 136.40 & 139.75 & 22.00 & 30.01 & 121.53 & 113.62 & 32.47 & 47.00 \\
5 & 119.12 & 118.34 & 16.74 & 27.22 & 114.59 & 112.42 & 20.42 & 22.58 \\
\bottomrule
\end{tabular}%
}

    \captionof{figure}{Network traffic received by iPerf for Experiments 1 to 5 - Last 5 minutes}
    \label{fig:boxplot-receive-baseline}
\end{figure}

Figure \ref{fig:boxplot-receive-baseline} displays a boxplot illustrating the distribution of data samples received by iPerf during the last 5 minutes of the experiment, highlighting the central tendency and variability of the data. In the first boxplot, corresponding to the first experiment, the central value of the rectangle represents the median, or second quartile (Q2), reflecting the central tendency of the samples at 154.04 Mbps. The lower and upper limits of the rectangle correspond to the first (Q1) and third (Q3) quartiles, defining the IQR, which covers 50\% of the samples around the median. A smaller IQR amplitude indicates less fluctuation in the sample data. The lines extending from the rectangle, known as whiskers, establish the limits for identifying outliers, represented by circles. For all the other experiments, two boxes are shown: one for the priority slice and another for the aggregate data of the other slices.

As illustrated in Figure \ref{fig:boxplot-receive-baseline}, a reduction in the data received by iPerf is observed, considering the average of the priority slice. In the first three experiments, the distribution of samples is similar when comparing the first and third quartiles, suggesting the availability of spare resources. However, the UPFs face processing difficulties as contention increases, resulting in a reduction in the average received data. In Experiments 4 and 5, a more significant decrease in Q1 and Q3 values is observed, indicating reduced availability of resources. In Experiment 5, the smaller IQR variation reflects system saturation, operating with all available resources continuously, limiting fluctuations in received bandwidth and bringing the system closer to its operational limit, minimizing significant variations.

\begin{figure}[!htbp]

\centering
  \includegraphics[width=0.9\linewidth]{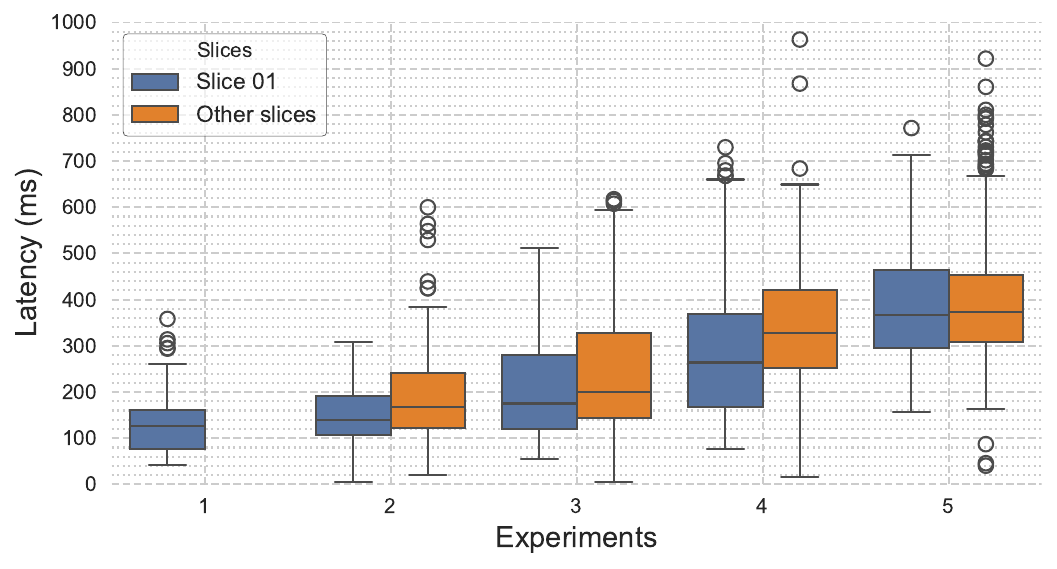}
    \resizebox{\columnwidth}{!}{%
    \begin{tabular}{c|cccc|cccc}
    \toprule
    \multirow{3}{*}{Experiment} & \multicolumn{4}{c}{\textcolor{slice1}{\rule{1em}{1em}} Slice 1} & \multicolumn{4}{|c}{\textcolor{others}{\rule{1em}{1em}} Other slices} \\ 
     & Mean & Median & Standard Deviation & IQR & Mean & Median & Standard Deviation & IQR\\
    & (ms) &  (ms) &  (ms)       & (ms) & (ms) &  (ms) &  (ms) &  (ms)\\
    \midrule
    1 & 132.72 & 125.80 & 66.92 & 84.75 &   &   &   &   \\
    2 & 151.35 & 139.40 & 63.58 & 84.50 & 198.88 & 166.80 & 112.36 & 119.20 \\
    3 & 209.14 & 175.30 & 114.06 & 160.10 & 245.57 & 200.00 & 141.56 &   \\
    4 & 289.89 & 263.90 & 149.28 & 199.80 & 340.47 & 328.40 & 136.60 &   \\
    5 & 388.90 & 367.75 & 122.85 & 168.20 & 393.03 & 372.35 & 129.86 & 145.20 \\
    \bottomrule
    \end{tabular}%
    }
  \captionof{figure}{Latency between UE and iPerf for experiments 1 to 5 - Last 5 minutes}
  \label{fig:boxplot-latency-baseline}
\end{figure}

Latency observed in each slice is represented in Figure \ref{fig:boxplot-latency-baseline}, where non-priority slices were grouped. In Experiment 1, the median latency is approximately 120 ms, with little dispersion, but some outliers above 300 ms. Similar to network traffic, progressive degradation occurs with each added slice, resulting in an increase in the median latency of Slice 1 and greater dispersion.

\FloatBarrier
\subsection{Experiments with CPU Limitation}\label{sec:exp-baseline-cpu}

To understand the behavior of traffic in slices under CPU limitation, experiments 6 to 8 were conducted. Table \ref{tab:04CPU} presents the imposed limitations for the slices, with the priority slice fixed at 1000 millicpu, a value more than sufficient for the priority slice. For the other slices, they have no restriction in experiment 6, are limited to 500 millicpu in experiment 7, and later to 250 millicpu in experiment 8. To ensure better control over CPU limitations, the Guaranteed QoS class of Kubernetes was adopted, where the CPU limit and request have identical values.

\begin{table}
    \begin{tabular}{lcccccc}
    \hline
    \multicolumn{1}{|l|}{} & \multicolumn{3}{c|}{\textbf{Slice 01}} & \multicolumn{3}{c|}{\textbf{Other Slices}} \\
    \multicolumn{1}{|l|}{\multirow{-2}{*}{}} & \multicolumn{1}{c|}{CPU} & \multicolumn{1}{c|}{Nice} & \multicolumn{1}{c|}{Bandwith} & \multicolumn{1}{c|}{CPU} & \multicolumn{1}{c|}{Nice} & \multicolumn{1}{c|}{Bandwith} \\ \hline
    \multicolumn{1}{|l|}{Experiment 06} & \multicolumn{1}{c|}{1000} & \multicolumn{1}{c|}{} & \multicolumn{1}{c|}{} & \multicolumn{1}{c|}{} & \multicolumn{1}{c|}{} & \multicolumn{1}{c|}{} \\ \hline
    \rowcolor[HTML]{C0C0C0} 
    \multicolumn{1}{|l|}{\cellcolor[HTML]{C0C0C0}Experiment 07} & \multicolumn{1}{c|}{\cellcolor[HTML]{C0C0C0}1000} & \multicolumn{1}{c|}{\cellcolor[HTML]{C0C0C0}} & \multicolumn{1}{c|}{\cellcolor[HTML]{C0C0C0}} & \multicolumn{1}{c|}{\cellcolor[HTML]{C0C0C0}500} & \multicolumn{1}{c|}{\cellcolor[HTML]{C0C0C0}} & \multicolumn{1}{c|}{\cellcolor[HTML]{C0C0C0}} \\ \hline
    \multicolumn{1}{|l|}{Experiment 08} & \multicolumn{1}{c|}{1000} & \multicolumn{1}{c|}{} & \multicolumn{1}{c|}{} & \multicolumn{1}{c|}{250} & \multicolumn{1}{c|}{} & \multicolumn{1}{c|}{} \\ \hline
    \end{tabular}
    \caption{Experiments with CPU limitation}
    \label{tab:04CPU}
\end{table}

\begin{figure}[H]
\centering
\includegraphics[width=0.9\linewidth]{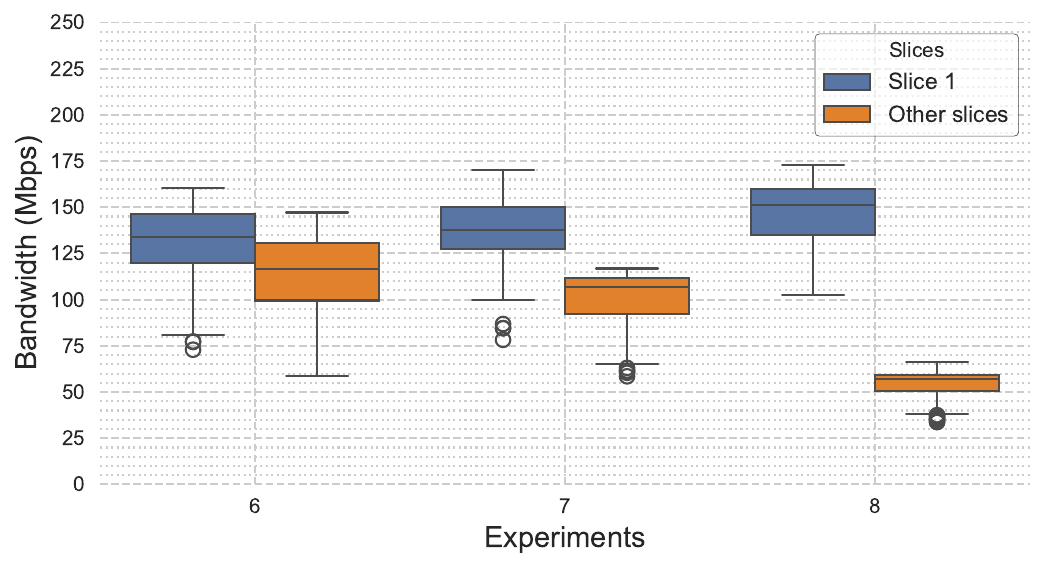}
\resizebox{\columnwidth}{!}{%
\begin{tabular}{c|cccc|cccc}
\toprule
\multirow{3}{*}{Experiment} & \multicolumn{4}{c}{\textcolor{slice1}{\rule{1em}{1em}} Slice 1} & \multicolumn{4}{|c}{\textcolor{others}{\rule{1em}{1em}} Other slices} \\
& Mean & Median & Std. Dev. & IQR & Mean & Median & Std. Dev. & IQR \\
& (Mbps) & (Mbps) & (Mbps) & (Mbps) & (Mbps) & (Mbps) & (Mbps) & (Mbps)\\
\midrule
6 & 130.49 & 133.92 & 20.32 & 26.08 & 114.17 & 116.45 & 19.07 & 30.97 \\
7 & 136.43 & 137.85 & 18.25 & 22.74 & 101.77 & 106.63 & 12.61 & 19.22 \\
8 & 147.09 & 151.44 & 16.29 & 24.88 & 54.98 & 57.21 & 6.20 & 8.65 \\
\bottomrule
\end{tabular}%
}
\caption{Network traffic received by iPerf for Experiments 6 to 8 - Last 5 minutes}

\label{fig:boxplot-receive-cpu}
\end{figure}

By analyzing the graph in Figure \ref{fig:boxplot-receive-cpu}, it can be observed that limiting the CPU for secondary slices helps the priority slice achieve higher throughput. The average bandwidth received by the iPerf servers in the priority slice in each experiment increases significantly as the CPU limit for the secondary slices is reduced. Initially, the average is 138 Mbps, increasing to 142 Mbps, and finally reaching 151 Mbps in experiments 6, 7, and 8, respectively. However, this configuration negatively impacts the other slices, resulting in a significant reduction in the bandwidth received by iPerf in Experiment 8, reaching almost half of the initial value (comparing to baseline Experiment 5).

\begin{figure}[!htbp]
\centering
\includegraphics[width=0.9\linewidth]{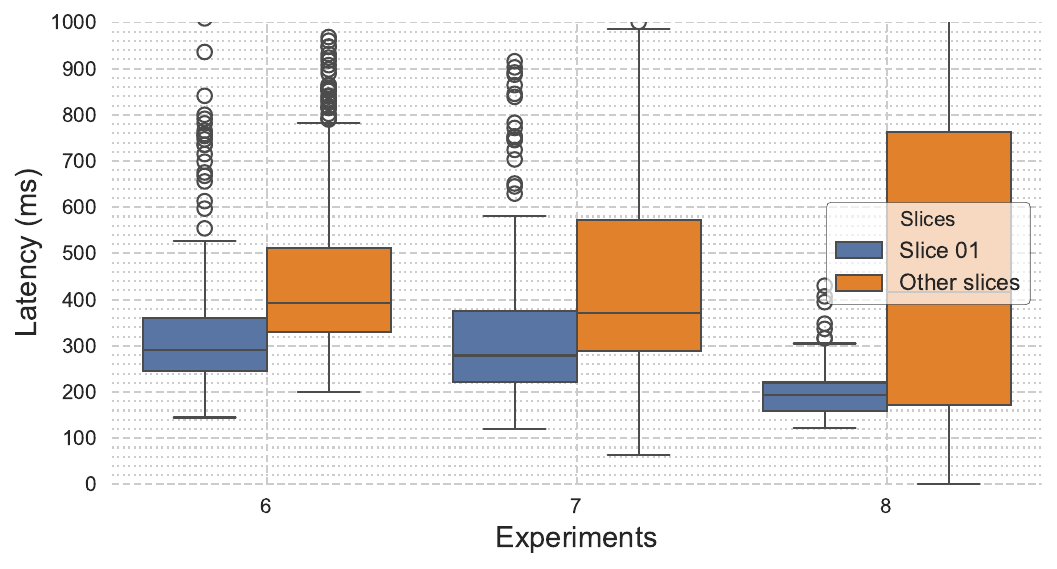}
\resizebox{\columnwidth}{!}{%
\begin{tabular}{c|cccc|cccc}
\toprule
\multirow{3}{*}{Experiment} & \multicolumn{4}{c}{\textcolor{slice1}{\rule{1em}{1em}} Slice 1} & \multicolumn{4}{|c}{\textcolor{others}{\rule{1em}{1em}} Other slices} \\
& Mean & Median & Std. Dev. & IQR & Mean & Median & Std. Dev. & IQR\\
& (ms) & (ms) & (ms) & (ms) & (ms) & (ms) & (ms) & (ms)\\
\midrule
6 & 375.06 & 291.20 & 215.81 & 114.65 & 471.08 & 392.50 & 245.37 & 181.67 \\
7 & 364.13 & 279.15 & 227.51 & 152.60 & 448.90 & 371.60 & 248.46 & \\
8 & 203.36 & 192.80 & 61.77 & 61.30 & 502.66 & 415.60 & 396.12 & \\
\bottomrule
\end{tabular}%
}
\captionof{figure}{Latency between UE and iPerf for Experiments 6 to 8 - Last 5 minutes}
\label{fig:boxplot-latency-cpu}
\end{figure}

The graph in Figure \ref{fig:boxplot-latency-cpu} shows the latency between the UEs and the iPerfs throughout the three experiments (6, 7, and 8) with CPU limitation. It is possible to observe that the priority slice consistently exhibited lower latency, while the other slices experienced higher latency with greater dispersion, suggesting inferior performance. This indicates that limiting the CPU for the other slices results in lower delays for the priority slice. The graph is capped at 1000 ms to facilitate comparison with the other graphs.

The graphs clearly demonstrate the effect of CPU limitation in a 5G network edge environment with slicing. The priority slice is able to maintain both low latency and high throughput in a stable manner, reflecting its privileged position in resource allocation. In contrast, the other slices face greater challenges, with higher latency and more inconsistent performance, particularly under severe CPU limitations, as seen in Experiment 8. This highlights the importance of proper CPU allocation to ensure Quality of Service in sliced 5G networks.

\FloatBarrier
\subsection{Experiments enabling CPU prioritization}\label{sec:exp-baseline-nice}

In experiments 9 and 10, CPU prioritization is performed on the slices using the Linux \textit{nice} command, applied to the UPF network function. This ensures it has higher priority in the CPU scheduler compared to other processes on the node. In this context, the value -5 is assigned to the prioritized slice, as shown in Table \ref{tab:04NICE}, while the value 5 is used for the other slices in Experiment 10. Note that, when no \textit{nice} value is specified to a process, the default value is zero.

\begin{table}[h]

\centering
\begin{tabular}{lcccccc}
\hline
\multicolumn{1}{|l|}{} & \multicolumn{3}{c|}{\textbf{Slice 01}} & \multicolumn{3}{c|}{\textbf{Other Slices}} \\
\multicolumn{1}{|l|}{\multirow{-2}{*}{}} & \multicolumn{1}{c|}{CPU} & \multicolumn{1}{c|}{Nice} & \multicolumn{1}{c|}{Bandwidth} & \multicolumn{1}{c|}{CPU} & \multicolumn{1}{c|}{Nice} & \multicolumn{1}{c|}{Bandwidth} \\ \hline

\rowcolor[HTML]{C0C0C0} 
\multicolumn{1}{|l|}{\cellcolor[HTML]{C0C0C0}Experiment 09} & \multicolumn{1}{c|}{\cellcolor[HTML]{C0C0C0}} & \multicolumn{1}{c|}{\cellcolor[HTML]{C0C0C0}-5} & \multicolumn{1}{c|}{\cellcolor[HTML]{C0C0C0}} & \multicolumn{1}{c|}{\cellcolor[HTML]{C0C0C0}} & \multicolumn{1}{c|}{\cellcolor[HTML]{C0C0C0}} & \multicolumn{1}{c|}{\cellcolor[HTML]{C0C0C0}} \\ \hline
\multicolumn{1}{|l|}{Experiment 10} & \multicolumn{1}{c|}{} & \multicolumn{1}{c|}{-5} & \multicolumn{1}{c|}{} & \multicolumn{1}{c|}{} & \multicolumn{1}{c|}{5} & \multicolumn{1}{c|}{} \\ \hline

\end{tabular}
\caption{Experiments with CPU prioritization}
\label{tab:04NICE}
\end{table}

\begin{figure}[!htbp]

\centering
  \includegraphics[width=0.9\linewidth]{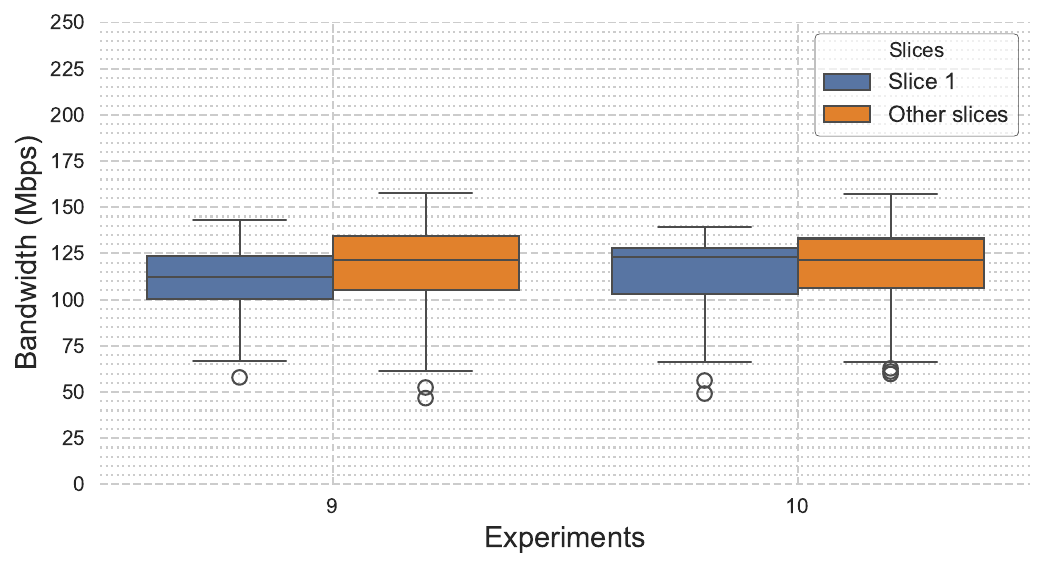}
\resizebox{\columnwidth}{!}{%
\begin{tabular}{c|cccc|cccc}
\toprule
\multirow{3}{*}{Experiment} & \multicolumn{4}{c}{\textcolor{slice1}{\rule{1em}{1em}} Slice 1} & \multicolumn{4}{|c}{\textcolor{others}{\rule{1em}{1em}} Other slices} \\ 
 & Mean & Median & Standard Deviation & IQR & Mean & Median & Standard Deviation & IQR\\
& (Mbps) &  (Mbps) &  (Mbps)       & (Mbps) & (Mbps) &  (Mbps) &  (Mbps) &  (Mbps)\\
\midrule
9 & 110.74 & 112.50 & 18.74 & 22.94 & 117.89 & 121.44 & 20.53 & 29.20 \\
10 & 114.18 & 123.26 & 19.93 & 24.89 & 118.85 & 121.23 & 19.07 & 26.73 \\
\bottomrule
\end{tabular}%
}
 \captionof{figure}{Network traffic received by iPerf for experiments 9 and 10 - Last 5 minutes}
  \label{fig:boxplot-receive-nice}
\end{figure}

The analysis of the chart in Figure \ref{fig:boxplot-receive-nice}, regarding the data received by the \textit{iPerfs} in the last 5 minutes of each experiment, shows that in Experiment 9, the traffic remains concentrated between 100 and 124 Mbps, while in Experiment 10, it ranges between 104 and 128 Mbps. This indicates that there is no significant difference between the two experiments, as their boundaries are close, indicating similar variability.

\begin{figure}[h]
  \centering
  \includegraphics[width=0.9\linewidth]{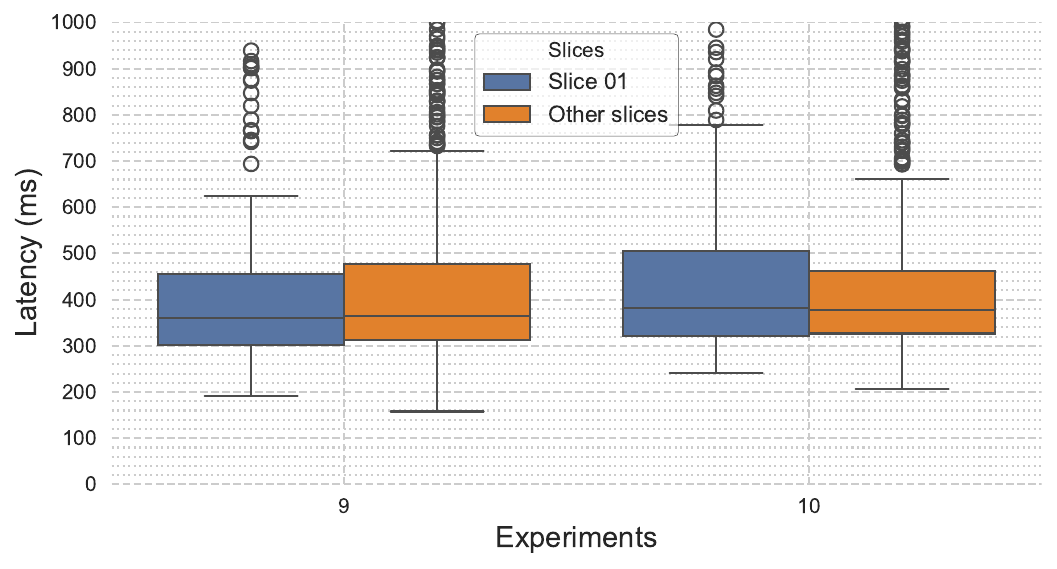}
  \resizebox{\columnwidth}{!}{%
    \begin{tabular}{c|cccc|cccc}
    \toprule
    \multirow{3}{*}{Experiment} & \multicolumn{4}{c}{\textcolor{slice1}{\rule{1em}{1em}} Slice 1} & \multicolumn{4}{|c}{\textcolor{others}{\rule{1em}{1em}} Other slices} \\ 
     & Mean & Median & Standard Deviation & IQR & Mean & Median & Standard Deviation & IQR\\
    & (ms) &  (ms) &  (ms)       & (ms) & (ms) &  (ms) &  (ms) &  (ms)\\
    \midrule
    9 & 440.69 & 359.45 & 209.90 & 155.10 & 432.84 & 365.05 & 199.31 & 165.30 \\
    10 & 466.85 & 382.15 & 228.54 & 183.05 & 458.38 & 377.10 & 223.66 & 135.35 \\
    \bottomrule
    \end{tabular}%
    }
  \captionof{figure}{Latency between experiments 9 and 10 - Last 5 minutes}
  \label{fig:boxplot-latency-nice}
\end{figure}

The latency represented in Figure \ref{fig:boxplot-latency-nice} exhibits similar behavior to that observed in baseline Experiment 5, suggesting that the \textit{nice} parameter does not affect latency. The median latency values are approximately 360 ms for any slice (prioritized or not).

Although CPU prioritization is implemented to benefit the prioritized slice, the results do not indicate significant advantages for it. Even with the prioritized \gls{upf} receiving more processing time on the CPU, which theoretically would allow greater efficiency in handling network packets, the observed performance is similar to that of the other slices. This behavior can be explained by the fact that the bottleneck lies in the network, which limits the availability of packets for processing by the network function.

\FloatBarrier
\subsection{Experiments enabling bandwidth limitation}\label{sec:exp-band}

In Experiments 11 to 13, the \texttt{egress-bandwidth} property of \textit{Kubernetes/Cilium} was applied to restrict the UPF's outgoing bandwidth, as detailed in Table \ref{tab:04BAND}. In experiment 11, the configuration was set to zero, activating the bandwidth control mechanism to enable the creation of priority queues without imposing a rate limit on the queues. This scenario allowed the analysis of the default behavior of bandwidth control without specific restrictions. Subsequently, the bandwidth was limited to 150 Mbps for non-priority slices in Experiment 12 and to 75 Mbps in Experiment 13. No restriction was imposed on the priority slice, as the intention was to maximize the use of available bandwidth.

\begin{table}[h]
\caption{Experiments with bandwidth limitation activation}
\label{tab:04BAND}
\centering
\begin{tabular}{lcccccc}
\hline
\multicolumn{1}{|l|}{} & \multicolumn{3}{c|}{\textbf{Slice 01}} & \multicolumn{3}{c|}{\textbf{Other Slices}} \\ 
\multicolumn{1}{|l|}{\multirow{-2}{*}{}} & \multicolumn{1}{c|}{CPU} & \multicolumn{1}{c|}{Nice} & \multicolumn{1}{c|}{Bandwidth} & \multicolumn{1}{c|}{CPU} & \multicolumn{1}{c|}{Nice} & \multicolumn{1}{c|}{Bandwidth} \\ \hline

\rowcolor[HTML]{C0C0C0} 
\multicolumn{1}{|l|}{\cellcolor[HTML]{C0C0C0}Experiment 11} & \multicolumn{1}{c|}{\cellcolor[HTML]{C0C0C0}} & \multicolumn{1}{c|}{\cellcolor[HTML]{C0C0C0}} & \multicolumn{1}{c|}{\cellcolor[HTML]{C0C0C0}0} & \multicolumn{1}{c|}{\cellcolor[HTML]{C0C0C0}} & \multicolumn{1}{c|}{\cellcolor[HTML]{C0C0C0}} & \multicolumn{1}{c|}{\cellcolor[HTML]{C0C0C0}0} \\ \hline
\multicolumn{1}{|l|}{Experiment 12} & \multicolumn{1}{c|}{} & \multicolumn{1}{c|}{} & \multicolumn{1}{c|}{} & \multicolumn{1}{c|}{} & \multicolumn{1}{c|}{} & \multicolumn{1}{c|}{150} \\ \hline
\rowcolor[HTML]{C0C0C0} 
\multicolumn{1}{|l|}{\cellcolor[HTML]{C0C0C0}Experiment 13} & \multicolumn{1}{c|}{\cellcolor[HTML]{C0C0C0}} & \multicolumn{1}{c|}{\cellcolor[HTML]{C0C0C0}} & \multicolumn{1}{c|}{\cellcolor[HTML]{C0C0C0}} & \multicolumn{1}{c|}{\cellcolor[HTML]{C0C0C0}} & \multicolumn{1}{c|}{\cellcolor[HTML]{C0C0C0}} & \multicolumn{1}{c|}{\cellcolor[HTML]{C0C0C0}75} \\ \hline

\end{tabular}
\centering
\end{table}

\begin{figure}[!h]

\centering
  \centering
  \includegraphics[width=0.9\linewidth]{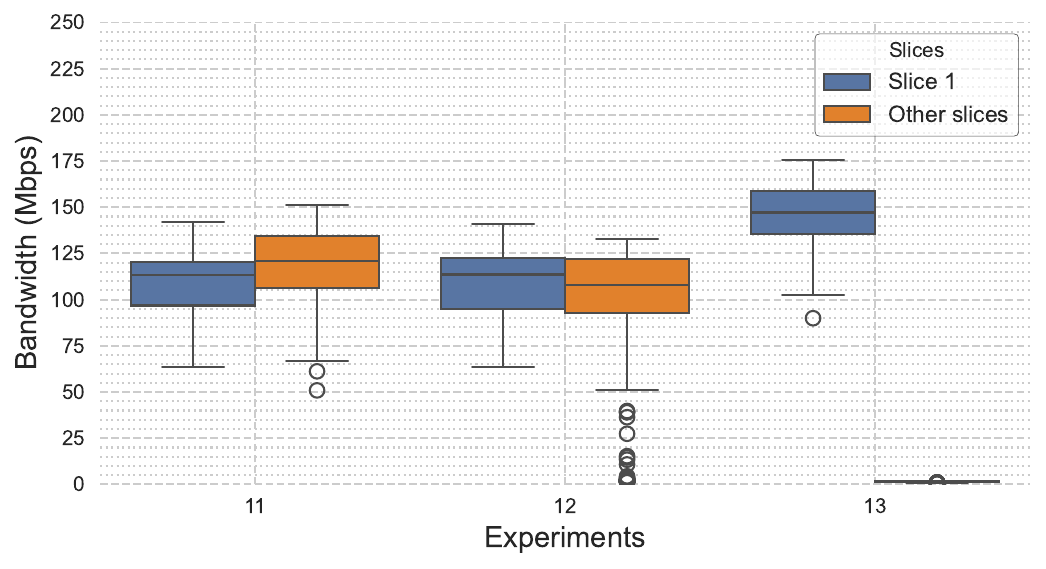}
\resizebox{\columnwidth}{!}{%
\begin{tabular}{c|cccc|cccc}
\toprule
\multirow{3}{*}{Experiment} & \multicolumn{4}{c}{\textcolor{slice1}{\rule{1em}{1em}} Slice 1} & \multicolumn{4}{|c}{\textcolor{others}{\rule{1em}{1em}} Other slices} \\ 
 & Mean & Median & Standard Deviation & IQR & Mean & Median & Standard Deviation & IQR\\
& (Mbps) &  (Mbps) &  (Mbps)       & (Mbps) & (Mbps) &  (Mbps) &  (Mbps) &  (Mbps)\\
\midrule
11 & 110.59 & 113.43 & 16.32 & 23.56 & 117.96 & 121.09 & 20.47 & 28.40 \\
12 & 107.71 & 113.62 & 17.14 & 27.69 & 101.39 & 108.12 & 30.13 & 29.06 \\
13 & 145.12 & 147.17 & 17.19 & 23.06 & 1.60 & 1.59 & 0.30 & 0.33 \\
\bottomrule
\end{tabular}%
}
\captionof{figure}{Network traffic received by iPerf for experiments 11 to 13 - Last 5 minutes}
\label{fig:boxplot-receive-band}
\end{figure}

Analyzing the chart in Figure \ref{fig:boxplot-receive-band}, which illustrates the data received by the iPerfs in the last 5 minutes of the experiments, it is observed that the bandwidth limitation contributes to a slight improvement in the priority slice in Experiments 11 and 12. However, with the imposition of a severe bandwidth limit in Experiment 13, a significant improvement in the performance of the priority slice is observed, while the secondary slices are negatively impacted, presenting an average of data received close to zero, rendering their operation unfeasible.

\begin{figure}[H]

  \centering
  \includegraphics[width=0.9\linewidth]{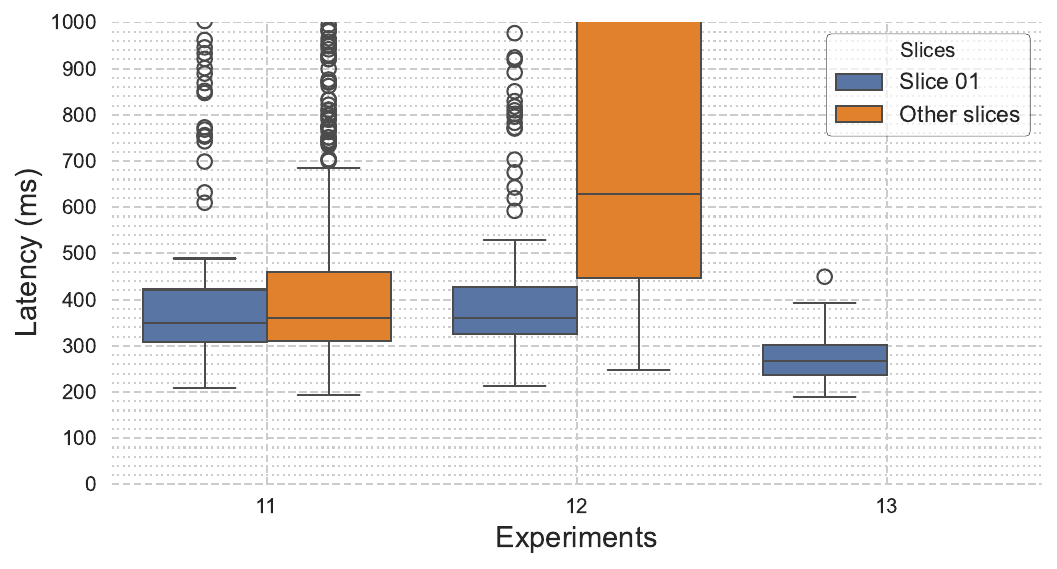}
\resizebox{\columnwidth}{!}{%
\begin{tabular}{c|cccc|cccc}
\toprule
\multirow{3}{*}{Experiment} & \multicolumn{4}{c}{\textcolor{slice1}{\rule{1em}{1em}} Slice 1} & \multicolumn{4}{|c}{\textcolor{others}{\rule{1em}{1em}} Other slices} \\ 
 & Mean & Median & Standard Deviation & IQR & Mean & Median & Standard Deviation & IQR\\
& (ms) &  (ms) &  (ms)       & (ms) & (ms) &  (ms) &  (ms) &  (ms)\\
\midrule
11 & 451.85 & 349.90 & 251.74 & 114.45 & 451.74 & 360.85 & 252.42 & 150.62 \\
12 & 434.29 & 360.55 & 199.53 & 101.85 & 6161.78 & 629.05 & 18016.42 & 732.42 \\
13 & 274.32 & 266.40 & 53.20 & 65.38 & 101343.93 & 100186.25 & 29914.92 & 40230.70 \\
\bottomrule
\end{tabular}%
} 
  \captionof{figure}{Latency between UE and iPerf for experiments 11 to 13 - Last 5 minutes}
  \label{fig:boxplot-latency-band}
\end{figure}

The latency represented in the chart of Figure \ref{fig:boxplot-latency-band} shows approximate values for Experiments 11 and 12, averaging around 440 ms. However, in Experiment 13, the latency decreases to approximately 270 ms, with almost no presence of outliers, in contrast to the other experiments, which exhibit several outliers. For the other slices, high latency variability is observed in Experiment 12, exceeding 1000 ms. The same occurs in Experiment 13, although it is not visible in the chart as it exceeds the limits established for the graphs.

Finally, the use of bandwidth limitation can be useful but should be applied with caution, as it tends to significantly increase the latency of low-priority slices. In the case of experiment 13, an extremely negative impact was observed on the secondary slices, presenting an atypical behavior compared to the other experiments. Memory usage dropped to almost zero in all secondary \gls{upf}s. Since the bandwidth limitation is applied at the UPF's outgoing traffic, it is plausible that packet drops occurred due to the exhaustion of some internal queue, allowing the service to remain active but with severely compromised performance.

\FloatBarrier
\subsection{Experiments with CPU limitation and prioritization}\label{sec:exp-cpu-nice}

In Experiments 14, 15, and 16, a combination of CPU limitation, applied independently in Experiments 6, 7, and 8, and CPU prioritization via \textit{nice}, implemented in Experiment 10, was used as described in Table \ref{tab:04CPUNICE}. For the priority slice, the CPU reservation was set to 1000 milicpu, and the \textit{nice} parameter was assigned a value of -5 in all experiments. For non-priority slices, the \textit{nice} parameter was set to 5, following the configuration of Experiment 10, while the CPU limitations followed the values specified in Experiments 6 (no limitation), 7 (500 milicpu), and 8 (250 milicpu).

\begin{table}[h]
\centering
\begin{tabular}{lcccccc}
\hline
\multicolumn{1}{|l|}{} & \multicolumn{3}{c|}{\textbf{Slice 01}} & \multicolumn{3}{c|}{\textbf{Other Slices}} \\  
\multicolumn{1}{|l|}{\multirow{-2}{*}{}} & \multicolumn{1}{c|}{CPU} & \multicolumn{1}{c|}{Nice} & \multicolumn{1}{c|}{Bandwidth} & \multicolumn{1}{c|}{CPU} & \multicolumn{1}{c|}{Nice} & \multicolumn{1}{c|}{Bandwidth} \\ \hline

\multicolumn{1}{|l|}{Experiment 14 (6 + 10)} & \multicolumn{1}{c|}{1000} & \multicolumn{1}{c|}{-5} & \multicolumn{1}{c|}{} & \multicolumn{1}{c|}{} & \multicolumn{1}{c|}{5} & \multicolumn{1}{c|}{} \\ \hline
\rowcolor[HTML]{C0C0C0} 
\multicolumn{1}{|l|}{\cellcolor[HTML]{C0C0C0}Experiment 15 (7 + 10)} & \multicolumn{1}{c|}{\cellcolor[HTML]{C0C0C0}1000} & \multicolumn{1}{c|}{\cellcolor[HTML]{C0C0C0}-5} & \multicolumn{1}{c|}{\cellcolor[HTML]{C0C0C0}} & \multicolumn{1}{c|}{\cellcolor[HTML]{C0C0C0}500} & \multicolumn{1}{c|}{\cellcolor[HTML]{C0C0C0}5} & \multicolumn{1}{c|}{\cellcolor[HTML]{C0C0C0}} \\ \hline
\multicolumn{1}{|l|}{Experiment 16 (8 + 10)} & \multicolumn{1}{c|}{1000} & \multicolumn{1}{c|}{-5} & \multicolumn{1}{c|}{} & \multicolumn{1}{c|}{250} & \multicolumn{1}{c|}{5} & \multicolumn{1}{c|}{} \\ \hline

\end{tabular}
\centering
\caption{Experiments with CPU limitation and prioritization activation}
\label{tab:04CPUNICE}

\end{table}

\begin{figure}[H]

\centering
  \includegraphics[width=0.9\linewidth]{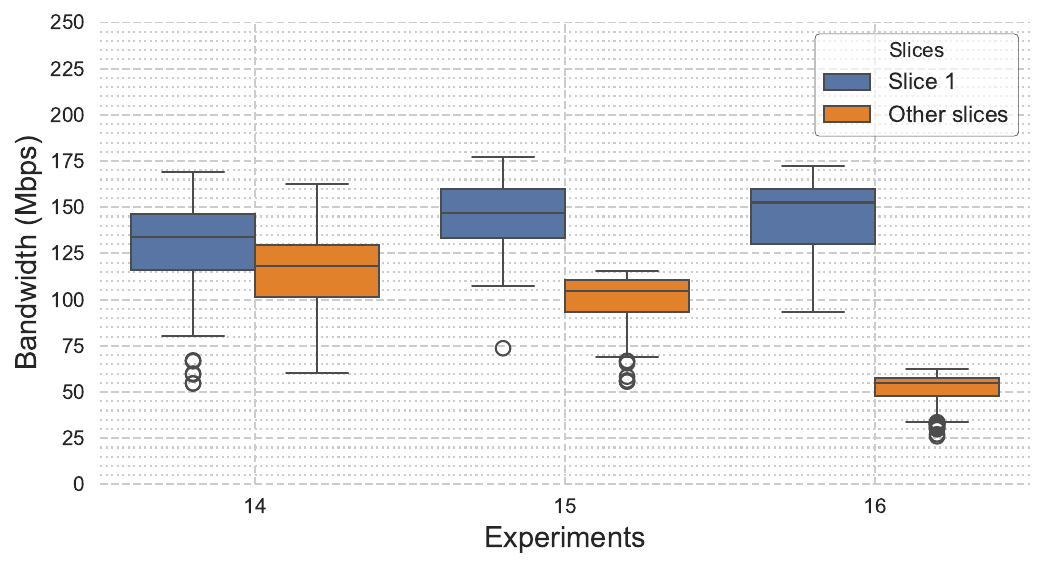}
\resizebox{\columnwidth}{!}{%
\begin{tabular}{c|cccc|cccc}
\toprule
\multirow{3}{*}{Experiment} & \multicolumn{4}{c}{\textcolor{slice1}{\rule{1em}{1em}} Slice 1} & \multicolumn{4}{|c}{\textcolor{others}{\rule{1em}{1em}} Other slices} \\ 
 & Mean & Median & Standard Deviation & IQR & Mean & Median & Standard Deviation & IQR\\
& (Mbps) &  (Mbps) &  (Mbps)       & (Mbps) & (Mbps) &  (Mbps) &  (Mbps) &  (Mbps)\\
\midrule
14 & 128.92 & 134.09 & 25.42 & 30.20 & 115.07 & 118.28 & 20.07 & 28.25 \\
15 & 146.52 & 146.78 & 17.33 & 26.71 & 101.26 & 104.66 & 11.68 & 17.53 \\
16 & 145.74 & 152.56 & 18.10 & 29.39 & 52.04 & 54.76 & 7.52 & 9.57 \\
\bottomrule
\end{tabular}%
}
\captionof{figure}{Network traffic received by iPerf for experiments 14 to 16 - Last 5 minutes}
\label{fig:boxplot-receive-cpu-nice}
\end{figure}

The chart in Figure \ref{fig:boxplot-receive-cpu-nice} shows the data received by the iPerf instances under the combination of CPU limitation and prioritization. The average data received by iPerf during the last 5 minutes of the experiment is 136, 150, and 151 Mbps for Experiments 14, 15, and 16, respectively. The combination demonstrates an improvement when comparing Experiment 15 (average of 146 Mbps) with Experiment 7 (average of 136 Mbps), where CPU limitation was applied in isolation. Additionally, the rate obtained in Experiment 15 is also better than that of Experiment 10, where only CPU prioritization was applied. This behavior suggests that a combination of resource restriction parameters can, in fact, be more effective to produce slice isolation effects than each individual mechanism.

\begin{figure}[h]

  \centering
  \includegraphics[width=0.9\linewidth]{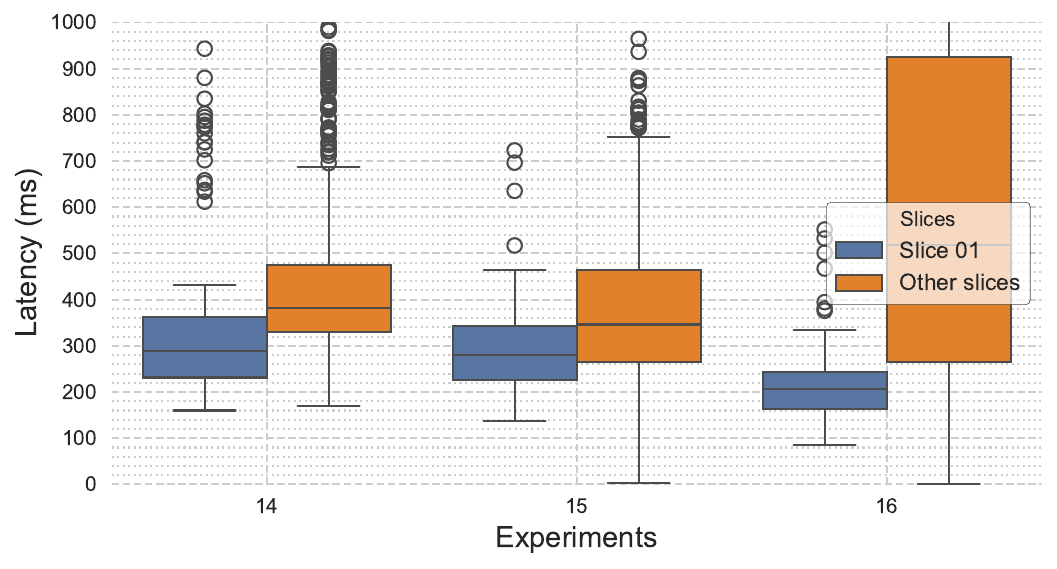}
\resizebox{\columnwidth}{!}{%
\begin{tabular}{c|cccc|cccc}
\toprule
\multirow{3}{*}{Experiment} & \multicolumn{4}{c}{\textcolor{slice1}{\rule{1em}{1em}} Slice 1} & \multicolumn{4}{|c}{\textcolor{others}{\rule{1em}{1em}} Other slices} \\ 
 & Mean & Median & Standard Deviation & IQR & Mean & Median & Standard Deviation & IQR\\
& (ms) &  (ms) &  (ms)       & (ms) & (ms) &  (ms) &  (ms) &  (ms)\\
\midrule
14 & 371.81 & 288.60 & 223.23 & 131.30 & 459.26 & 382.10 & 214.52 & 143.38 \\
15 & 294.66 & 280.00 & 99.49 & 115.80 & 383.69 & 346.30 & 184.31 & 197.52 \\
16 & 219.14 & 206.80 & 88.94 & 81.20 & 654.85 & 518.10 & 506.43 & 660.40 \\
\bottomrule
\end{tabular}%
} 
\captionof{figure}{Latency between UE and iPerf for experiments 14 to 16 - Last 5 minutes}
\label{fig:boxplot-latency-cpu-nice}
\end{figure}

The chart in Figure \ref{fig:boxplot-latency-cpu-nice} shows the latency between the \gls{ue}s and iPerf during the last 5 minutes, comparing the experiments that combine CPU limitation with CPU prioritization. When compared to the experiments presented in the chart in Figure \ref{fig:boxplot-latency-cpu}, the values are similar, indicating no significant difference between the experiments.

In general, the combinations of resource control mechanisimos used in Experiments 15 and 16 result in an improvement in throughput. The use of the \textit{nice} parameter, which did not show significant effects in previous experiments when used in isolation, now appears to contribute to slice isolation when combined with CPU limitation.

\FloatBarrier
\subsection{Experiments with CPU limitation and bandwidth limitation}\label{sec:exp-cpu-band}

In Experiments 17 to 22, a combination of isolated CPU limitation experiments (Experiments 6 to 8) and bandwidth limitation experiments (Experiments 11 and 12) was performed. The combination with Experiment 13 was not conducted due to its atypical behavior caused by limiting the bandwidth to 75 Mbps of low priority slices. Table \ref{tab:04CPUBAND} details the combined CPU and network bandwidth configurations used in Experiments 17 to 22. In all experiments, the priority slice maintained a constant configuration of 1000m CPU. In contrast, the other slices had their CPU configurations adjusted according to the experiment: in Experiment 17, there were no CPU limits; in Experiment 18, the limits were set at 500m; in Experiment 19, they were reduced to 250m; in Experiment 20, the limits were again removed; and in Experiments 21 and 22, the limits were re-established at 500m and 250m, respectively.

Regarding bandwidth control, in Experiments 17, 18, and 19, the control was activated without a specific limit for the non-priority slices, as was done in Experiment 11. In Experiments 20 to 22, a limit of 150 Mbps was set for the bandwidth of non-priority slices, as occurs in Experiment 12. Meanwhile, the priority slice remained unrestricted. These configurations were designed to examine the impact of different CPU and bandwidth limits on the performance of the priority slice compared to the other slices in the environment.

\begin{figure}[h]

  \centering
  \includegraphics[width=0.9\linewidth]{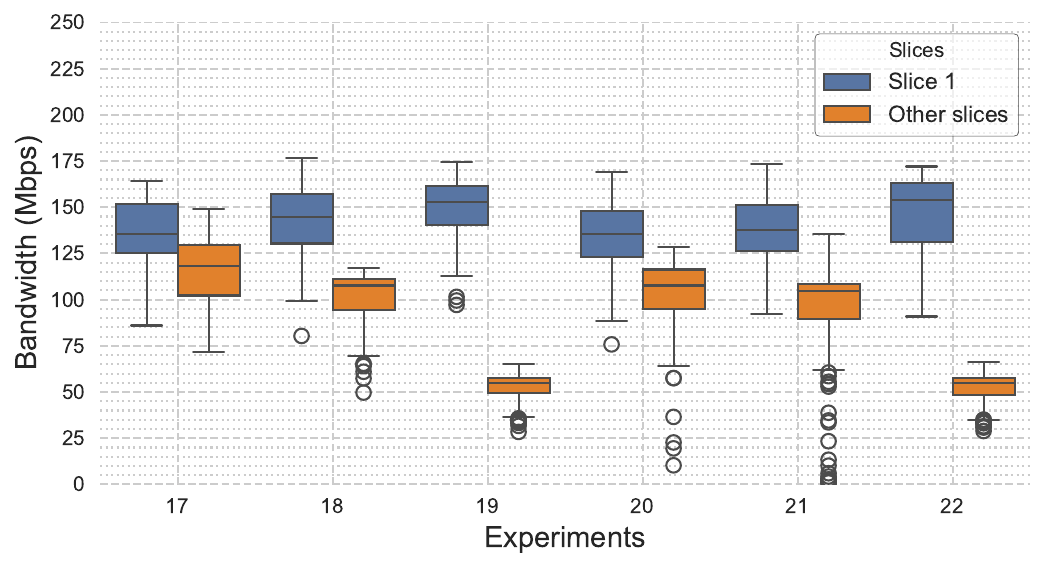}
\resizebox{\columnwidth}{!}{%
\begin{tabular}{c|cccc|cccc}
\toprule
\multirow{3}{*}{Experiment} & \multicolumn{4}{c}{\textcolor{slice1}{\rule{1em}{1em}} Slice 1} & \multicolumn{4}{|c}{\textcolor{others}{\rule{1em}{1em}} Other slices} \\ 
 & Mean & Median & Standard Deviation & IQR & Mean & Median & Standard Deviation & IQR\\
& (Mbps) &  (Mbps) &  (Mbps)       & (Mbps) & (Mbps) &  (Mbps) &  (Mbps) &  (Mbps)\\
\midrule
17 & 135.16 & 135.30 & 18.39 & 26.40 & 115.50 & 118.23 & 17.59 & 27.30 \\
18 & 143.11 & 144.89 & 17.90 & 26.82 & 101.95 & 107.67 & 11.87 & 16.87 \\
19 & 149.08 & 152.73 & 16.59 & 20.79 & 53.15 & 54.96 & 6.60 & 8.55 \\
20 & 132.82 & 135.51 & 19.42 & 25.03 & 103.76 & 107.68 & 16.18 & 21.50 \\
21 & 136.75 & 137.50 & 17.73 & 24.86 & 97.82 & 104.84 & 23.66 & 18.60 \\
22 & 145.20 & 153.78 & 21.93 & 32.12 & 53.05 & 55.13 & 6.51 & 8.96 \\
\bottomrule
\end{tabular}%
}
\caption{Network traffic received by iPerf for experiments 17 to 22 - Last 5 minutes}
\label{fig:boxplot-receive-cpu-band}
\end{figure}

Analyzing Figure \ref{fig:boxplot-receive-cpu-band}, on average, Experiment 17 shows an improvement when compared to Experiment 20, in which a bandwidth limitation of 150 Mbps was added for the other slices. The same occurs with Experiments 18 and 21, and 19 and 22. This indicates that limiting the bandwidth of the other slices also negatively affects the priority slice. The number of outliers is also higher when the bandwidth limitation is included.

\begin{table}[h]

\centering
\begin{tabular}{lcccccc}
\hline
\multicolumn{1}{|l|}{} & \multicolumn{3}{c|}{\textbf{Slice 01}} & \multicolumn{3}{c|}{\textbf{Other Slices}} \\  
\multicolumn{1}{|l|}{\multirow{-2}{*}{}} & \multicolumn{1}{c|}{CPU} & \multicolumn{1}{c|}{Nice} & \multicolumn{1}{c|}{Bandwidth} & \multicolumn{1}{c|}{CPU} & \multicolumn{1}{c|}{Nice} & \multicolumn{1}{c|}{Bandwidth} \\ \hline

\rowcolor[HTML]{C0C0C0} 
\multicolumn{1}{|l|}{\cellcolor[HTML]{C0C0C0}Experiment 17 (6 + 11)} & \multicolumn{1}{c|}{\cellcolor[HTML]{C0C0C0}1000} & \multicolumn{1}{c|}{\cellcolor[HTML]{C0C0C0}} & \multicolumn{1}{c|}{\cellcolor[HTML]{C0C0C0}0} & \multicolumn{1}{c|}{\cellcolor[HTML]{C0C0C0}} & \multicolumn{1}{c|}{\cellcolor[HTML]{C0C0C0}} & \multicolumn{1}{c|}{\cellcolor[HTML]{C0C0C0}0} \\ \hline
\multicolumn{1}{|l|}{Experiment 18 (7 + 11)} & \multicolumn{1}{c|}{1000} & \multicolumn{1}{c|}{} & \multicolumn{1}{c|}{0} & \multicolumn{1}{c|}{500} & \multicolumn{1}{c|}{} & \multicolumn{1}{c|}{0} \\ \hline
\rowcolor[HTML]{C0C0C0} 
\multicolumn{1}{|l|}{\cellcolor[HTML]{C0C0C0}Experiment 19 (8 + 11)} & \multicolumn{1}{c|}{\cellcolor[HTML]{C0C0C0}1000} & \multicolumn{1}{c|}{\cellcolor[HTML]{C0C0C0}} & \multicolumn{1}{c|}{\cellcolor[HTML]{C0C0C0}0} & \multicolumn{1}{c|}{\cellcolor[HTML]{C0C0C0}250} & \multicolumn{1}{c|}{\cellcolor[HTML]{C0C0C0}} & \multicolumn{1}{c|}{\cellcolor[HTML]{C0C0C0}0} \\ \hline
\multicolumn{1}{|l|}{Experiment 20 (6 + 12)} & \multicolumn{1}{c|}{1000} & \multicolumn{1}{c|}{} & \multicolumn{1}{c|}{} & \multicolumn{1}{c|}{} & \multicolumn{1}{c|}{} & \multicolumn{1}{c|}{150} \\ \hline
\rowcolor[HTML]{C0C0C0} 
\multicolumn{1}{|l|}{\cellcolor[HTML]{C0C0C0}Experiment 21 (7 + 12)} & \multicolumn{1}{c|}{\cellcolor[HTML]{C0C0C0}1000} & \multicolumn{1}{c|}{\cellcolor[HTML]{C0C0C0}} & \multicolumn{1}{c|}{\cellcolor[HTML]{C0C0C0}} & \multicolumn{1}{c|}{\cellcolor[HTML]{C0C0C0}500} & \multicolumn{1}{c|}{\cellcolor[HTML]{C0C0C0}} & \multicolumn{1}{c|}{\cellcolor[HTML]{C0C0C0}150} \\ \hline
\multicolumn{1}{|l|}{Experiment 22 (8 + 12)} & \multicolumn{1}{c|}{1000} & \multicolumn{1}{c|}{} & \multicolumn{1}{c|}{} & \multicolumn{1}{c|}{250} & \multicolumn{1}{c|}{} & \multicolumn{1}{c|}{150} \\ \hline

\end{tabular}
\centering
\caption{Experiments with CPU and bandwidth limitation activation}
\label{tab:04CPUBAND}
\end{table}

\begin{figure}[h]

\centering
  \centering
  \includegraphics[width=0.9\linewidth]{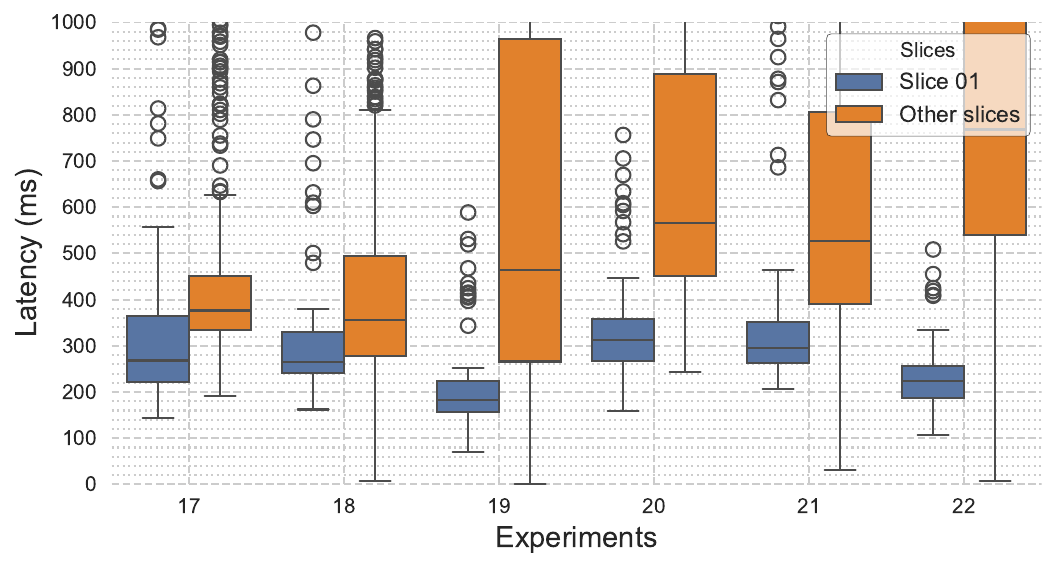}
\label{tab:cpu-band-receive}
\resizebox{\columnwidth}{!}{%
\begin{tabular}{c|cccc|cccc}
\toprule
\multirow{3}{*}{Experiment} & \multicolumn{4}{c}{\textcolor{slice1}{\rule{1em}{1em}} Slice 1} & \multicolumn{4}{|c}{\textcolor{others}{\rule{1em}{1em}} Other slices} \\ 
 & Mean & Median & Standard Deviation & IQR & Mean & Median & Standard Deviation & IQR\\
& (ms) &  (ms) &  (ms)       & (ms) & (ms) &  (ms) &  (ms) &  (ms)\\
\midrule
17 & 346.82 & 268.30 & 224.72 & 142.60 & 434.19 & 376.50 & 176.33 & 117.10 \\
18 & 324.93 & 265.80 & 179.37 & 89.05 & 428.34 & 356.10 & 234.03 & 214.93 \\
19 & 211.76 & 183.00 & 109.15 & 67.40 & 687.96 & 465.10 & 596.11 & 698.12 \\
20 & 335.00 & 312.90 & 120.49 & 90.70 & 1025.81 & 566.05 & 1905.67 & 438.55 \\
21 & 369.64 & 296.30 & 214.15 & 88.65 & 5247.41 & 527.10 & 20928.62 & 416.40 \\
22 & 230.73 & 224.15 & 73.26 & 69.68 & 981.35 & 768.35 & 685.47 & 839.65 \\
\bottomrule
\end{tabular}%
} 
  
  \captionof{figure}{Latency between UE and iPerf for experiments 17 to 22 - Last 5 minutes}
  \label{fig:boxplot-latency-cpu-band}
\end{figure}

The chart in Figure \ref{fig:boxplot-latency-cpu-band} presents the latency between the UE and iPerf during the last 5 minutes for Experiments 17 to 22. On average, Experiment 17 exhibits lower latency compared to Experiment 20, with similar trends observed between experiments 18 and 21, and 19 and 22. Introducing bandwidth limitations for the non-priority slices tends to increase latency variability, as shown by the greater number of outliers in the experiments where the bandwidth limitation was applied.

\FloatBarrier
\subsection{Experiments with CPU limitation, bandwidth limitation, and prioritization}\label{sec:04CPUBANDNICE}

Experiments 23 to 26 combine the three mechanisms previously used. Table \ref{tab:04CPUBANDNICE} details these combinations and the configurations of CPU, \textit{nice}-based prioritization, and bandwidth control. In all cases, the priority slice (Slice 1) was configured with the prioritization used in Experiment 10, that is, with \textit{nice} set to -5, while the other slices were set to a value of 5. The CPU limit follows the configurations from Experiments 6 and 7, with a CPU limit of 1000 milicpu for all priority slices. CPU limits of 500 milicpu were applied in Experiments 25 and 26. Regarding bandwidth limitation, in Experiments 24 and 26, a bandwidth limit of 150 Mbps was assigned, while in Experiments 25 and 26, a CPU limit of 500 milicpu was defined for the non-priority slices.

\begin{table}[h]

\centering
\begin{tabular}{lcccccc}
\hline
\multicolumn{1}{|l|}{} & \multicolumn{3}{c|}{\textbf{Slice 01}} & \multicolumn{3}{c|}{\textbf{Other Slices}} \\
\multicolumn{1}{|l|}{\multirow{-2}{*}{}} & \multicolumn{1}{c|}{CPU} & \multicolumn{1}{c|}{Nice} & \multicolumn{1}{c|}{Bandwidth} & \multicolumn{1}{c|}{CPU} & \multicolumn{1}{c|}{Nice} & \multicolumn{1}{c|}{Bandwidth} \\ \hline

\rowcolor[HTML]{C0C0C0} 
\multicolumn{1}{|l|}{\cellcolor[HTML]{C0C0C0}Experiment 23 (6 + 10 + 11)} & \multicolumn{1}{c|}{\cellcolor[HTML]{C0C0C0}1000} & \multicolumn{1}{c|}{\cellcolor[HTML]{C0C0C0}-5} & \multicolumn{1}{c|}{\cellcolor[HTML]{C0C0C0}0} & \multicolumn{1}{c|}{\cellcolor[HTML]{C0C0C0}} & \multicolumn{1}{c|}{\cellcolor[HTML]{C0C0C0}5} & \multicolumn{1}{c|}{\cellcolor[HTML]{C0C0C0}0} \\ \hline
\multicolumn{1}{|l|}{Experiment 24 (6 + 10 + 12)} & \multicolumn{1}{c|}{1000} & \multicolumn{1}{c|}{-5} & \multicolumn{1}{c|}{} & \multicolumn{1}{c|}{} & \multicolumn{1}{c|}{5} & \multicolumn{1}{c|}{150} \\ \hline
\rowcolor[HTML]{C0C0C0} 
\multicolumn{1}{|l|}{\cellcolor[HTML]{C0C0C0}Experiment 25 (7 + 10 + 11)} & \multicolumn{1}{c|}{\cellcolor[HTML]{C0C0C0}1000} & \multicolumn{1}{c|}{\cellcolor[HTML]{C0C0C0}-5} & \multicolumn{1}{c|}{\cellcolor[HTML]{C0C0C0}0} & \multicolumn{1}{c|}{\cellcolor[HTML]{C0C0C0}500} & \multicolumn{1}{c|}{\cellcolor[HTML]{C0C0C0}5} & \multicolumn{1}{c|}{\cellcolor[HTML]{C0C0C0}0} \\ \hline
\multicolumn{1}{|l|}{Experiment 26 (7 + 10 + 12)} & \multicolumn{1}{c|}{1000} & \multicolumn{1}{c|}{-5} & \multicolumn{1}{c|}{} & \multicolumn{1}{c|}{500} & \multicolumn{1}{c|}{5} & \multicolumn{1}{c|}{150} \\ \hline

\end{tabular}
\centering
\caption{Experiments combining CPU limitation, prioritization, and bandwidth limitation}
\label{tab:04CPUBANDNICE}
\end{table}

\begin{figure}[h]

\centering
  \includegraphics[width=0.9\linewidth]{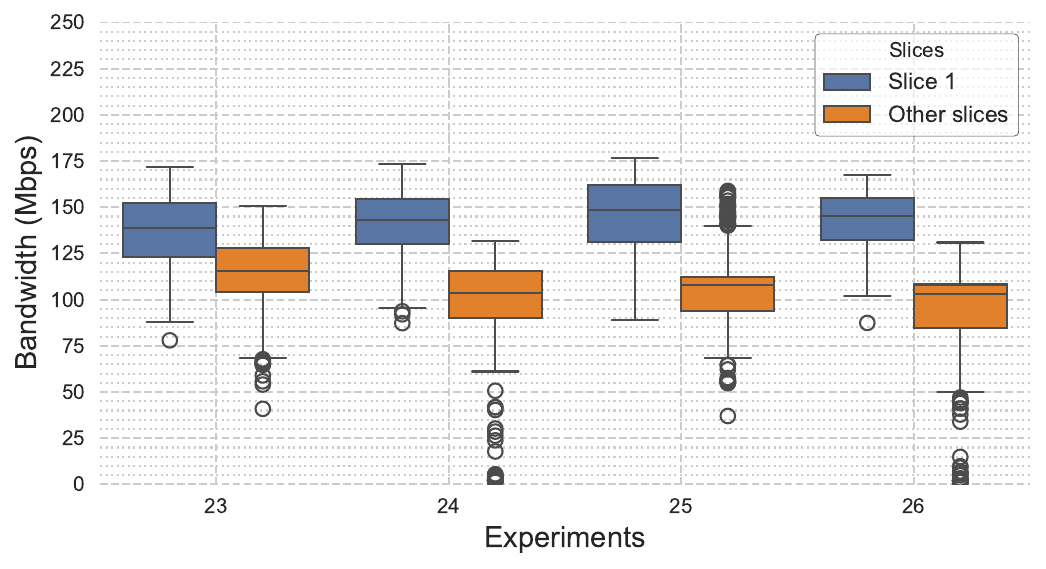}
\resizebox{\columnwidth}{!}{%
\begin{tabular}{c|cccc|cccc}
\toprule
\multirow{3}{*}{Experiment} & \multicolumn{4}{c}{\textcolor{slice1}{\rule{1em}{1em}} Slice 1} & \multicolumn{4}{|c}{\textcolor{others}{\rule{1em}{1em}} Other slices} \\ 
 & Mean & Median & Standard Deviation & IQR & Mean & Median & Standard Deviation & IQR\\
& (Mbps) &  (Mbps) &  (Mbps)       & (Mbps) & (Mbps) &  (Mbps) &  (Mbps) &  (Mbps)\\
\midrule
23 & 137.50 & 138.97 & 20.02 & 29.31 & 113.51 & 115.48 & 19.40 & 24.13 \\
24 & 140.40 & 142.86 & 19.69 & 24.38 & 96.72 & 103.49 & 28.80 & 25.58 \\
25 & 145.83 & 148.62 & 20.04 & 30.78 & 107.05 & 108.16 & 20.56 & 18.46 \\
26 & 142.14 & 145.40 & 17.19 & 22.45 & 89.66 & 103.29 & 30.36 & 23.71 \\
\bottomrule
\end{tabular}%
}
  \captionof{figure}{Network traffic received by iPerf for experiments 23 to 26 - Last 5 minutes}
  \label{fig:boxplot-receive-cpu-nice-band}
\end{figure}

The chart in Figure \ref{fig:boxplot-receive-cpu-nice-band} presents a comparison of the data received via \textit{iPerf} in Experiments 23 to 26. It can be observed that the priority slice remains close to 149 Mbps, with a significant improvement in Experiment 25, where the median reaches 152 Mbps. In Experiment 26, the median of the received data is 150 Mbps, slightly lower than in Experiment 25; however, it stands out for having the lowest standard deviation among the experiments, with a value of 9.37.

\begin{figure}[h]

  \centering
  \includegraphics[width=0.9\linewidth]{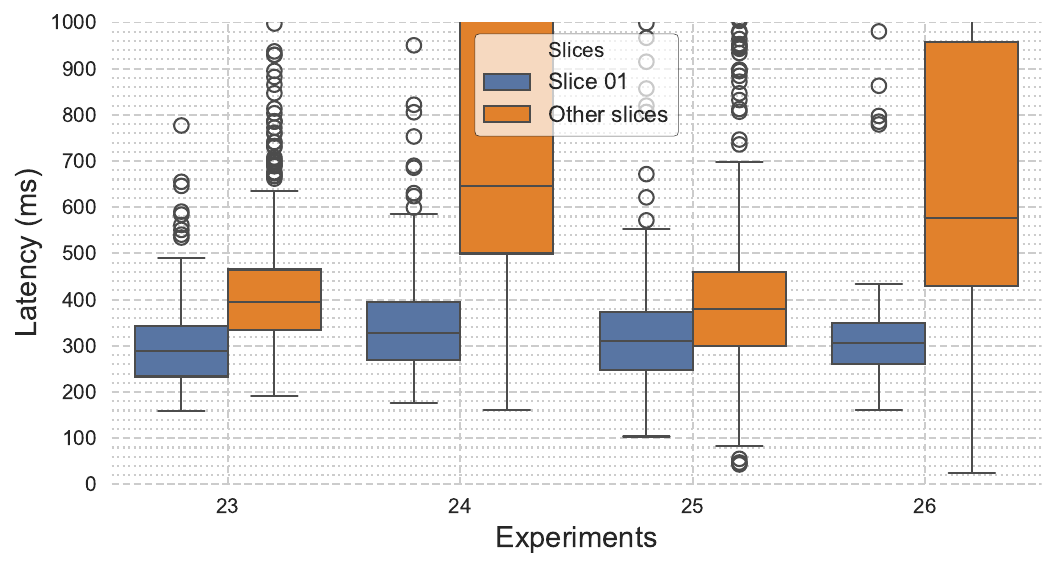}
\resizebox{\columnwidth}{!}{%
\begin{tabular}{c|cccc|cccc}
\toprule
\multirow{3}{*}{Experiment} & \multicolumn{4}{c}{\textcolor{slice1}{\rule{1em}{1em}} Slice 1} & \multicolumn{4}{|c}{\textcolor{others}{\rule{1em}{1em}} Other slices} \\ 
 & Mean & Median & Standard Deviation & IQR & Mean & Median & Standard Deviation & IQR\\
& (ms) &  (ms) &  (ms)       & (ms) & (ms) &  (ms) &  (ms) &  (ms)\\
\midrule
23 & 311.73 & 288.15 & 121.69 & 109.68 & 424.14 & 394.95 & 143.46 & 130.38 \\
24 & 405.54 & 327.05 & 266.59 & 126.38 & 4732.93 & 645.90 & 14391.46 & 672.80 \\
25 & 355.00 & 310.40 & 191.60 & 125.50 & 421.04 & 379.25 & 217.95 & 160.62 \\
26 & 380.70 & 305.60 & 241.49 & 88.20 & 6927.28 & 577.00 & 20900.67 & 528.25 \\
\bottomrule
\end{tabular}%
} 
  \captionof{figure}{Latency between UE and iPerf for experiments 23 to 26 - Last 5 minutes}
  \label{fig:boxplot-latency-cpu-nice-band}
\end{figure}

Examining Figure \ref{fig:boxplot-latency-cpu-nice-band}, which evaluates the latency between the \gls{ue} and \textit{iPerf}, it is noted that the priority slice maintains consistent values. However, the implementation of bandwidth limitation in Experiments 24 and 26 for non-priority slices results in a higher median latency compared to Experiments 23 and 25, as well as exhibiting greater variance.

\FloatBarrier
\subsection{General Discussion on Results} \label{sec:exp-discussion}

The evaluation of the experiments reveals that most of them exhibited a reduction in the average traffic received by \textit{iPerfs} for the priority slice greater than 10\% (or below 138.64 Mbps) when compared to Experiment 1 (best-case reference, with only the priority slice). Conversely, experiments 15, 18, 24, 25, and 26 showed a reduction of less than 10\% relative to this experiment. One can conclude that the parameters used for the resource control mechanisms in this set of experiments have the most positive effect in achieving the goal of isolating traffic for the priority slice and avoiding interference.

\begin{figure}[h]

  \centering
  \includegraphics[width=0.9\linewidth]{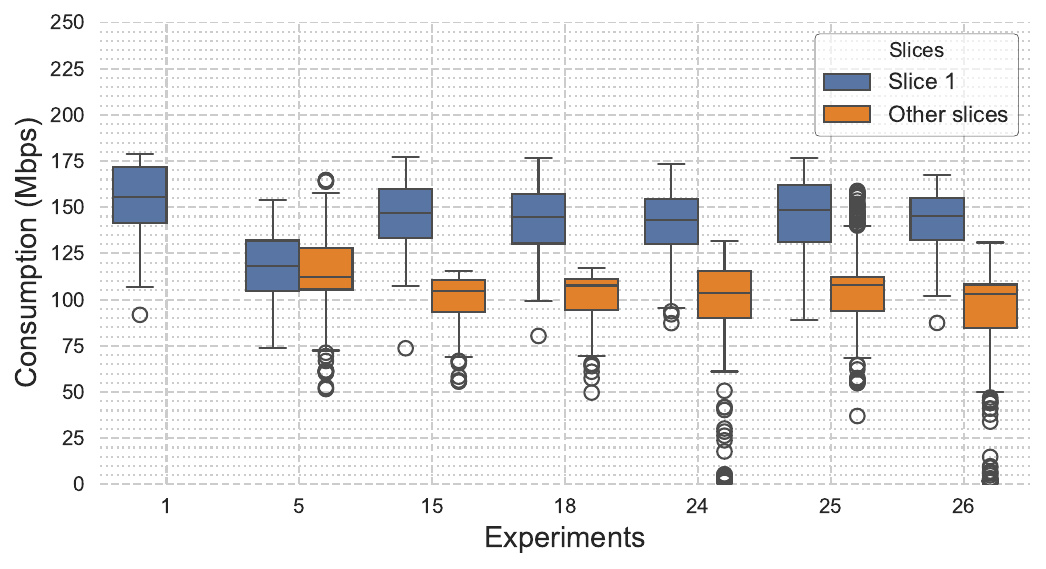}
\resizebox{0.8\columnwidth}{!}{%
\begin{tabular*}{\linewidth}{@{\extracolsep{\fill}}  c|cc|cc @{}}
\toprule
\multirow{2}{*}{Experiment} & \multicolumn{2}{c|}{\textcolor{slice1}{\rule{1em}{1em}} Slice 1 (Average)} & \multicolumn{2}{c}{\textcolor{others}{\rule{1em}{1em}} Other slices (Average)} \\ 
& (Mbps) &  (\%) &  (Mbps) &  (\%) \\
\midrule
1 & 154.04 &    &   &   \\
5 & 119.12 & 22.67 \% &  114.59 & \\
15 & 146.52 & 4.88 \% &  101.26 & 11.63 \% \\
18 & 143.11 & 7.10 \% &  101.95 & 11.03 \% \\
24 & 140.40 & 8.85 \% &  96.72  & 15.59 \% \\
25 & 145.83 & 5.33 \% &  107.05 & 6.58 \% \\
26 & 142.14 & 7.73 \% &  89.66  & 21.76 \% \\
\bottomrule
\end{tabular*}%
}

\caption{Network traffic received by iPerf for selected experiments - Last 5 minutes}
\label{fig:boxplot-receive-best}
\end{figure}

It is observed that experiments 8, 13, 16, 19, and 22, although showing less than a 10\% degradation for the priority slice, exhibited a reduction greater than 25\% for the other slices compared to Experiment 5 (worst-case reference, 5 slices without isolation). Figure \ref{fig:boxplot-receive-best}, developed to highlight the best experiments, illustrates that Experiment 15 shows a 4.88\% reduction for the priority slice, but achieves 11.63\% for the others. Experiment 25, in turn, demonstrates a 5.33\% reduction, with a difference of 0.45\%, and offers superior performance for the other slices, with a reduction of 6.58\%. The remaining experiments showed a reduction greater than 7\% in traffic for the priority slice and greater than 11\% for the other slices, with Experiment 26 showing a significant worsening to low-priority slices compared to the others.

The network traffic performance for the priority slice and the other slices highlights Experiment 25 as superior to the other experiments. The combined use of mechanisms emerges as crucial for effective slice isolation. Restricting CPU resources can promote a more equitable distribution, contributing to network stability, provided it is adjusted to service requirements. Additionally, greater CPU restrictions for lower-priority slices can benefit higher-priority ones. However, CPU prioritization proves effective only when combined with other mechanisms, being insufficient on its own. Bandwidth limitation should be applied with caution, as restricting to 150 Mbps already results in significant degradation for lower-priority slices, including the presence of several outliers.

\begin{figure}[h]

\centering
  \centering
  \includegraphics[width=0.9\linewidth]{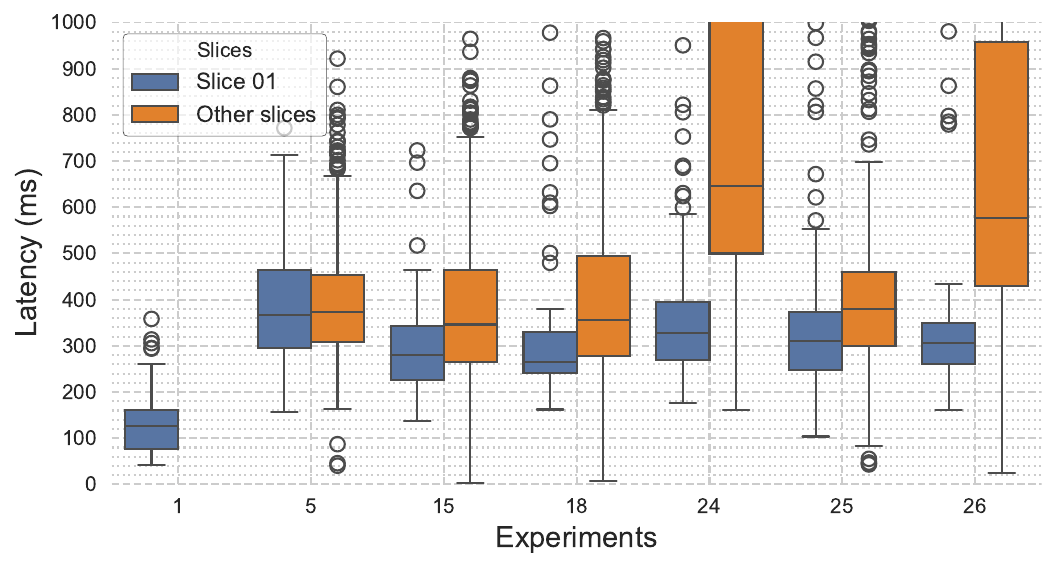}
\resizebox{0.8\columnwidth}{!}{%
\begin{tabular*}{\linewidth}{@{\extracolsep{\fill}}  c|cc|cc @{}}
\toprule
\multirow{2}{*}{Experiment} & \multicolumn{2}{c|}{\textcolor{slice1}{\rule{1em}{1em}} Slice 1 (Average)} & \multicolumn{2}{c}{\textcolor{others}{\rule{1em}{1em}} Other slices (Average)} \\ 
& (ms) &  (\%) &  (ms) &  (\%) \\
\midrule
1 & 154.04 &    &   &   \\
5  & 388.90 & 193.02 \% &  114.59 & \\
15 & 294.66 & 122.02 \% &  383.69 & 2.38 \% \\
18 & 324.93 & 144.82 \% &  428.34 & 8.98 \% \\
24 & 405.54 & 205.56 \% &  4732.93  & 1104.22 \% \\
25 & 355.00 & 167.48 \% &  421.04 & 7.13 \% \\
26 & 380.70 & 186.84 \% &  6927.28  & 1662.53 \% \\
\bottomrule
\end{tabular*}%
}
  \captionof{figure}{Latency between UE and iPerf for selected experiments - Last 5 minutes}
  \label{fig:boxplot-latency-best}
\end{figure}

Figure \ref{fig:boxplot-latency-best} illustrates the latency comparison for the same set of selected experiments, highlighting that the average latency for the priority slice is more than double compared to Experiment 1 (best-case scenario). Experiment 15 records the lowest average latency, below 300 ms, and shows latency for the other slices similar to that observed in Experiment 5 (worst-case scenario). In contrast, Experiments 24 and 26 exhibit substantially higher average latencies, reaching 4 and 6 seconds, respectively. Both use the 150 Mbps bandwidth limitation, suggesting that such a restriction exacerbates latency. Furthermore, the disparity between the mean and median is notable, indicating the presence of outliers that raise the mean.

Finally, Experiments 15 and 25 are the best, on average, for network isolation. If the goal is latency reduction, Experiment 15 has an advantage. Considering network throughput, Experiment 25 displays better results. The combination of CPU limitation and CPU prioritization has a positive effect, whereas bandwidth limitation seems to be more effective when merely activating the mechanism without applying a specific restriction.

\FloatBarrier
\section{Conclusion}\label{sec5}

This work discussed how 5G networks introduce significant advancements that expand opportunities for the development of advanced applications. A central aspect of this technology is network slicing, which enables traffic management according to the specific requirements of each application. Additionally, the use of open-source implementations of the 5G core not only facilitates the creation of private networks with limited resources, but also encourages the development of new technological solutions. However, the immaturity of these implementations still poses challenges, such as achieving efficient network traffic isolation. In this context, this work proposed a methodology that leverages native mechanisms of a Kubernetes-based edge/cloud infrastructure to enhance network slice isolation.

In this methodology, the mechanisms are applied to the User Plane Function (\gls{upf}), responsible for implementing network policies such as traffic redirection, quality of service, and individual data rate limitations for user equipment (\gls{ue}). However, applying these policies alone does not guarantee the complete isolation of a network slice. To address this issue, the proposed methodology suggests the use of mechanisms such as CPU limitation and prioritization, as well as bandwidth limitation at the \gls{upf} output, to minimize interference between different network slices.

To validate the effectiveness of the resource control mechanisms through the proposed methodology, a study environment was developed within a Kubernetes cluster. This choice is justified by Kubernetes' ability to provide advanced monitoring tools, such as Prometheus, which enable the collection of concrete data on network behavior and individual application performance. Furthermore, the developed and publicly available scripts allow for the reproduction of experiments in different environments, facilitating comparisons with future studies.

The experiments conducted in a hypothetical scenario within the Kubernetes cluster provided valuable insights into network traffic behavior and isolation in this environment. The analysis of the relationship between CPU allocation in a pod and network performance highlighted the importance of this factor in the results obtained, offering a series of valuable insights for ensuring adequate network slice isolation. The initial testing and evaluation model serves as a starting point for the development of a robust testing tool, enabling more comprehensive analyses in the context of 5G networks and facilitating significant advancements in the understanding and improvement of this technology.

The results of the experiments demonstrated that the combined use of CPU limitation and prioritization provides better performance, with a variation of approximately 5\% in data traffic compared to the traffic generated by a single network slice (best-case scenario, without competition for resources). However, it was observed that latency was negatively impacted in all experiments, highlighting opportunities for future studies focused on reducing the effects on latency.

It has been shown that the approach to network isolation using the available resources can be advantageous for private 5G networks, especially in scenarios with financial constraints. In summary, the experiments provided a detailed view of network behavior within a Kubernetes cluster and laid the groundwork for future investigations and developments, contributing to the continuous advancement of networks, with an eye on future generations and their practical applications.

For future work, it is recommended to explore new approaches to network isolation, such as comparing \gls{cni}s. Currently, only Cilium has been utilized, without fully exploring its potential due to the need to update the kernel on the nodes. This study could yield significant improvements by leveraging bandwidth limitation and traffic prioritization, which, unlike modifications to CPU limits, does not require application restarts. This characteristic is particularly interesting for applying artificial intelligence to dynamically adjust resource control mechanisms based on current network demands.






\backmatter





\bmhead{Acknowledgments}

This study was partially funded by CAPES Finance Code 001. We have also received support from two research projects funded by FAPESP jointly with MCTI, namely ``\textit{SFI2: Slicing Future Internet Infrastructures}'' (grant 18/23097-3) and ``\textit{PORVIR-5G: Programability, Orchestration and Virtualization in 5G Networks}'' (grant 20/05182-3).

\section{Data Availability}

The simulation codes and data files used in this research are accessible upon request.










\begin{appendices}






\end{appendices}


\bibliography{sn-bibliography}

\end{document}